\newcommand{\no}[1]{}
\newcommand{\bm}[1]{\mbox{\boldmath$#1$\unboldmath}}
\documentclass[12pt,amsmath,amssymb]{revtex4}
\usepackage{graphicx}
\begin{document}

\title{Low-Prandtl-number B\'enard-Marangoni convection in a
vertical magnetic field 
}


\author{Thomas Boeck      
}


\affiliation{Fakult\"at f\"ur Maschinenbau, Technische Universit\"at Ilmenau,\\
Postfach 100565, 98684 Ilmenau, Germany
}



\begin{abstract}
The effect of a homogeneous magnetic field
on surface-tension-driven B\'{e}nard convection is studied  
by means of 
direct numerical simulations. 
The flow is computed in a 
rectangular domain with periodic horizontal 
boundary conditions and the free-slip condition
on the bottom wall using a pseudospectral
Fourier-Chebyshev discretization. 
Deformations of the free
surface are neglected.
Two- and three-dimensional flows are computed
for either vanishing or 
small Prandtl number, which are 
typical of liquid metals.
The main focus of the paper is on a qualitative comparison of the
flow states  with the  non-magnetic case,
and on the effects associated with the possible
near-cancellation of the nonlinear and pressure terms
in the momentum equations for two-dimensional rolls.
In the three-dimensional case, the transition from a stationary 
hexagonal pattern at the onset of convection to  three-dimensional 
time-dependent convection is explored by a series of simulations
at zero Prandtl number.

\end{abstract}
\maketitle

\section{Introduction}
\label{intro}

B\'enard-Marangoni 
convection (BMC) in a plane fluid 
layer with an open surface
is the surface-tension-driven analog
to buoyancy-driven Rayleigh-B\'enard
convection (RBC) in a layer with rigid top and bottom
walls. Both systems present classic examples 
of hydrodynamic instabilities, and 
are generic models for convective
flows driven by buoyancy and Marangoni forces,
respectively.

Many works on BMC are devoted to the
regular cellular patterns as dissipative
structures, which are observed in highly
viscous fluids such as silicone oils, 
e.g. \cite{bestehorn1993,Golovin:1997,VanHook:1997,Eckert:1998}. 
In these experimental and theoretical investigations,
BMC is either stationary or  only weakly 
time-dependent, and the Prandtl number  $P$ is
large compared with unity
because of the  high kinematic viscosity.

Marangoni convection in low Prandtl number
fluids, i.e.  
liquid metals or semiconductor melts (with $P$ of order $10^{-2}$), 
plays an important role in industrial 
processes  such as crystal growth \cite{Davis:1987} or electron beam 
evaporation \cite{Pumir:Blumenfeld:1996}. 
Theoretical and experimental studies as well as numerical
simulations have
therefore been largely focused on configurations resembling  
actual industrial setups such as the floating
zone  \cite{Kuhlmann:Rath:1993,Levenstam:Amberg:1995}.

By contrast, BMC in liquid metals
has received comparatively little attention. 
The only experimental work  known to the author is the 
study of Ginde et al. \cite{Ginde:1989}, where liquid tin
was used. 
Theoretical studies of Dauby et al.  \cite{Dauby:1993} and 
of Thess \& Bestehorn \cite{Thess:Bestehorn:1995}
predict the occurrence of ``inverted" hexagons at the
onset of convection for sufficiently small Prandtl numbers. 
The orientation of the flow in such hexagonal cells is
opposite to that found at high Prandtl numbers.

Numerical studies of two- and three-dimensional
BMC were undertaken by the present author in collaboration 
with  A. Thess \cite{Boeck:1997,Boeck:1998,Boeck:1999}.
In \cite{Boeck:1997,Boeck:1999}, partiulcar attention was paid 
to the limit of vanishing Prandtl number. Earlier studies
of RBC had already explored low Prandtl numbers, and noted 
interesting behavior of two-dimensional roll solutions,
termed inertial convection \cite{Proctor:1977,Busse:Clever:1981}.
The analog of inertial convection in RBC
was observed and analyzed  in \cite{Boeck:1997} for BMC.
This type of convection is characterized by the dominance of
inertia over viscous forces  and a low energy dissipation. 
However, the two-dimensional rolls  are susceptible to
three-dimensional perturbations, and inertial convection is
therefore typically not realized in three-dimensional simulations 
\cite{Thual:1992,Boeck:1999}. Some indications for the existence of
inertial convection come from RBC experiments 
with  mercury and sodium \cite{Chiffaudel:1987,Kek:1993}, but those
are only based on integral heat flux measurements for a narrow range 
of Prandtl numbers.

In contrast to the non-conducting high-Prandtl-number liquids, the
B\'enard-Marangoni instability in liquid metals can be influenced by 
magnetic fields. For a vertical applied  field 
the linear stability theory has been thoroughly
explored by several authors 
\cite{Nield:1966,Wilson:1993,Wilson:1994,Thess:Nitschke:1995,Hashim:Wilson:1999}. 
In this configuration the magnetic field
does not introduce horizontal anisotropy. The magnetic field
suppresses the instability, i.e., the necessary temperature difference for
convective instability grows with the magnetic induction $B$.
For sufficiently strong $B$, the vertical structure of the
unstable mode exhibits Hartmann layers at the bottom and the free surface,
and the corresponding critical wavelength shrinks \cite{Thess:Nitschke:1995}.

A horizontal  component of the magnetic field introduces horizontal anisotropy,
but would  not suppress the instability because the roll axis of the normal mode 
can always be aligned with  the field. In this case no Lorentz force is produced.
The resulting anisotropy of the flow pattern has been discussed in
the context of linear theory 
\cite{Thess:Nitschke:1992}, but not on the basis of nonlinear
simulations.

The nonlinear behavior with a vertical magnetic field
has so far only been examined using
perturbation methods, which only apply near the onset of convection
\cite{Miladinova:2001}. The present paper continues the investigations
of nonlinear BMC with a vertical magnetic field using the direct 
simulation approach of \cite{Boeck:1997,Boeck:1999}, whereby large Marangoni
numbers and complex flows can be realized numerically. 
Both two- and three-dimensional simulations will be performed in order to
analyze the qualitative effects of the magnetic field on the particular
solutions described in \cite{Boeck:1997,Boeck:1999}. The choice
of parameters and boundary conditions is therefore guided by these
previous works.

The paper is organized as follows. In the following section
the basic equations and the numerical method will be described.
After that, the two-dimensional  and three-dimensional
simulation results are presented and analyzed in separate sections.
The paper ends with some conclusions and suggestions for future work.

\section{Basic equations and numerical method}
\label{sec:1}

The system under study is a planar liquid layer of thickness $d$ with 
a free upper surface, which is heated from below. The liquid 
has a finite electric conductivity $\sigma$. Buoyancy
and surface deflections are neglected.  
The isothermal bottom of the layer is located  at $z=0$
and the free surface at $z=d$.
Periodic boundary conditions apply in the  
$x$- and $y$-directions. The  thickness $d$
is chosen as lengthscale for non-dimensionalization.
The dimensionless quantities $L_x$ and $L_y$ denote 
the periodicity length with respect to $x$ and $y$.

As in \cite{Boeck:1997}, free-slip  conditions are assumed at the
bottom of the layer in order to compare with the non-magnetic case.
The heat flux density at the free surface is assumed fixed in
the present work, 
which corresponds to the limit of zero Biot number.
It is convenient to introduce  the deviation $\theta$ from the 
the conductive  temperature profile by writing
\begin{equation}
\label{condprofile}
T  = \theta +T_{b} - \Delta T_{0} z/d.
\end{equation}
Here $T_b$ is the prescribed bottom temperature and
$\Delta T_{0}$ the conductive temperature difference.
Notice that $ \theta(x,y,z) \le \Delta T_0 z/d$ since the fluid cannot
become hotter than the bottom wall.  
 The shear stresses at the
free surface are 
\begin{equation}
\label{marangoni1}
\rho \nu \frac{\partial \bm{v}}
{\partial z}=\nabla \Sigma =-\gamma\nabla T , 
\end{equation}
where the
coefficient  $\gamma$ characterizes the linear decrease in 
surface tension $\Sigma$ with temperature.  The symbols
$\bm{v}$,  $\rho$ and $\nu$ are the fluid velocity, density and 
kinematic viscosity. The vectors in eq. (\ref{marangoni1})
are understood to be projected onto the free surface. 

The Lorentz force and the induced current  
due to the external homogeneous magnetic field $B$
will be treated in the limit of low magnetic Reynolds number. 
As explained in \cite{Davidson:2001}, this approximation is justified
for length scales and velocities realized in the laboratory and
in industrial applications. It takes into account
the action of the magnetic field on the velocity field but ignores
the  reaction of  the flow on the magnetic field. 
The densities of
induced current and Lorentz force in this approximation are given by
\begin{equation}
\bm{j}= \sigma\left(
-\nabla \phi  + \bm{v}\times \bm{B}\right), \hskip3mm
\bm{f}=\bm{j}\times \bm{B}.
\end{equation}
The divergence of the induced current density $\bm{j}$ vanishes 
because of the extremly short charge relaxation time in comparison
with the time scales of magnetohydrodynamic flows 
\cite[Section 2.2]{Davidson:2001}. For this reason, the 
electric potential $\phi$ satisfies
\begin{equation}
 \nabla^2\phi= \nabla\cdot\left(\bm{v} \times \bm{B}\right).
\end{equation}
It is also assumed that the liquid is bounded by electrical insulators, i.e.
the normal component $j_z$ of the current density vanishes at the top
and the bottom of the layer. Using Ohm's law, $j_z=0$ translates into
boundary conditions on $\phi$.

The equations will be given in a form based on 
viscous velocity scaling, i.e. $\nu/d$ is the velocity
scale and $d^2/\nu$ the time scale. 
Furthermore,  $\nu \Delta T_0/\kappa$ serves as  
scale of $\theta$, where $\kappa$ denotes the
thermal diffusivity of the fluid. The magnetic field is oriented in
along the $\bm{e}_z$ unit vector, and the scale of the electric potential
is given by $B\nu$.
The dimensionless equations are
\begin{eqnarray}
\label{nsviscous}
\frac{\partial \bm{v}}{\partial t} +
(\bm{v}\cdot \nabla)\bm{v} &=& -\nabla p
+  \nabla^2 \bm{v} + Ha^2 \left(-\nabla \phi +\bm{v}\times \bm{e}_z\right)\times \bm{e}_z,\\
 \nabla \cdot\bm{v}  &=&  0,\\
\nabla^2 \phi &=&\nabla \cdot\left(\bm{v} \times \bm{e}_z\right),\\
\label{heatviscous}
 P\left( \frac{\partial \theta}{\partial t} +
(\bm{v}\cdot \nabla)\theta \right) &=& 
 \nabla^2  \theta + v_z. \no{heatviscous}
\end{eqnarray}
At the bottom $z=0$, the conditions
\begin{equation}
\label{bcbotfreeviscous}
\frac{\partial v_x}{\partial z}=\frac{\partial v_y}
{\partial z}=v_z=
 \theta=\frac{\partial \phi}{\partial z}=0\no{bcbotfreeviscous}
\end{equation}
apply. The boundary conditions at the top surface $z=1$ are 
\begin{equation}
\label{bctopviscous}
\frac{\partial v_x}{\partial z} + Ma\, \frac{\partial
\theta}{\partial x}=\frac{\partial v_y}{\partial z} + Ma\, \frac{\partial
\theta}{\partial y}= v_z=\frac{\partial \theta}{\partial
z}=\frac{\partial \phi}{\partial z}=0 \no{bctopviscous}
\end{equation}
with the Marangoni number 
\begin{equation}
Ma= \frac{\gamma \Delta T_0 d}{\rho \nu \kappa} 
\end{equation}
as control parameter for the forcing. The  remaining parameters
are the Prandtl number $P=\nu/\kappa$ and the   Hartmann number 
\begin{equation}
Ha= B d\sqrt{\frac{\sigma}{\rho \nu}}.
\end{equation}

The scale for $\theta$ has been chosen in such a way that 
a coupling of velocity and temperature is maintained also in the
limit $P=0$, i.e. 
when the left hand side in  (\ref{heatviscous}) vanishes.
This approximation is called the  (viscous)
zero--Prandtl--number limit.
The same limit has been considered in the work of Thual \cite{Thual:1992}
for RBC. Other  limiting cases based on 
different temperature and velocity
scalings are also discussed in this work,
and it is argued that these other cases either do not provide a uniform
approximation throughout the fluid domain, or model
transients or flows where thermal convection is 
merely a side effect.

Convective heat transport is
measured by the nondimensional Nusselt number $Nu$
defined as the ratio between the total heat flux and
the conductive heat flux. In contrast to convection between
isothermal plates with fixed temperatures,  convection reduces
the temperature drop across the layer. Since the contribution
of heat conduction to the total heat transport is thereby reduced,
one has to compute the conductive heat flux
for the present, reduced surface temperature.
The result reads
\begin{equation}
\label{newnusselt}
Nu = \frac{1 }{
1 - P\langle \theta_s  \rangle},\no{newnusselt}
\end{equation}
where
\begin{equation}
\langle \theta_s \rangle = \frac{1}{L_x L_y} \int_0^{L_x} \int_0^{L_y} 
\theta (x,y,1) d\, x \, d\, y
\end{equation}
denotes the mean perturbation of the surface temperature.

In the  zero--Prandtl--number limit 
$Nu$ equals unity. Results for $P>0$ can be extrapolated from 
$P=0$  if one regards the solution of the 
zero--Prandtl--number equations  as the leading term in 
an expansion in $P$.
The calculations given in \cite{Boeck:1997} lead to 
the following results for mean  surface temperature perturbation
$\langle \theta_s \rangle$ 
and $Nu-1$:
\begin{eqnarray}
\langle \theta_s \rangle &= &
P \int_0^1 \langle v_z\theta\rangle dz  + O(P^2),\\
\label{nusseltexpression}
Nu - 1 & = & P^2 \int_0^1 \langle v_z
 \theta\rangle dz  + O(P^3).
\end{eqnarray}
In these equations,  $v_z$ and $\theta$ are the solutions of the 
zero--Prandtl--number equations.
Due to its connection with the Nusselt number the quantity
$\overline{v_z\theta}$ (where the overbar symbol denotes the
volume average)
will be referred to as reduced Nusselt number.

Finally, the alternative scaling of the basic equations
based on the units $\kappa/d$ for  velocity, $\kappa B$ for the electric potential
and $\Delta T_0$ for the
temperature perturbation $\theta$ will be needed for later discussion.
The equations (\ref{nsviscous}-\ref{heatviscous}) then take the form
\begin{eqnarray}
\label{nsthermal}
\frac{\partial \bm{v}}{\partial t} +
(\bm{v}\cdot \nabla)\bm{v} &=& -\nabla p
+  P\left[\nabla^2 \bm{v} + Ha^2 \left(-\nabla \phi +\bm{v}\times \bm{e}_z\right)\times \bm{e}_z\right],\\
 \nabla \cdot\bm{v}  &=&  0,\\
\nabla^2 \phi &=&\nabla \cdot\left(\bm{v} \times \bm{e}_z\right),\\
\label{heatthermal}
 \frac{\partial \theta}{\partial t} +
(\bm{v}\cdot \nabla)\theta  &=& 
 \nabla^2  \theta + v_z. 
\end{eqnarray}
The boundary conditions (\ref{bcbotfreeviscous},\ref{bctopviscous}) remain
 unchanged, but for the Nusselt number one has to use 
\begin{equation}
\label{newnusselt-thermal}
Nu = \frac{1 }{
1 -\langle \theta_s  \rangle}
\end{equation}
instead of (\ref{newnusselt}).

The numerical code for two- and three-dimensional direct simulations
is based on equations (\ref{nsviscous}-\ref{heatviscous})
and a pseudospectral Fourier-Chebyshev discretization. It has been
described in \cite{Boeck:1999} for the
non-magnetic case. The modifications for the present case with
vertical magnetic field are analogous to the implementation of the 
electric potential and Lorentz force terms for the channel code discussed in 
\cite{Krasnov:2004}.

\section{Linear stability results}
The linear stability of the basic conductive state has been
treated in several papers, including asymptotic limits
of large  Hartmann number and Biot numbers as well as zero
and finite crispation numbers
\cite{Nield:1966,Wilson:1993,Wilson:1994,Thess:Nitschke:1995,Hashim:Wilson:1999}.
For the purposes of the
present paper the critical Marangoni and wavenumbers
for free-slip boundary conditions at the bottom are
required as a function of the Hartmann number. 

The velocity and temperature perturbations  for the
neutral stability problem are 
\begin{equation}
\label{stabilityansatz}
v_z(x,z)=W(z)\exp\left(ikx\right),\hskip5mm
\theta(x,z)=\Theta(z)\exp\left(ikx\right),
\end{equation}
whereby one obtains the stability equations 
\begin{equation}
\label{stabilityequations}
\left[\left(D^2-k^2\right)^2+
Ha^2 D^2\right] W(z)=0,\hskip5mm
\left[D^2-k^2\right]\Theta(z)=-W(z)
\end{equation}
and the boundary conditions
\begin{equation}
\label{stabilitybcs}
W(0)=W(1)=\Theta(0)=D\Theta(1)=D^2W(0)=D^2W(1) +Ma\, k^2\, \Theta(1)=0.
\end{equation}
The symbol $D$ denotes the $z$-derivative. 
Notice that the Prandtl number is absent from these equations since the
 neutral conditions are independent of $P$. Because of the linearity
 of the system (\ref{stabilityequations}-\ref{stabilitybcs}),
one can choose the normalization $D^2 W(1)=1$  to solve 
these equations without the Marangoni boundary condition
\begin{equation}
\label{ma-crit-eq}
Ma\, k^2\, \Theta(1)=- D^2W(1).
\end{equation} 
Eq. (\ref{ma-crit-eq}) 
 is then used to calculate the  Marangoni number $Ma(k)$ 
for  neutral conditions  at the chosen wavenumber $k$.
The solution of the fourth-order equation for $W$ and the second-order
equation for $\Theta$ is obtained simultaenously by a Chebyshev collocation
method.

\begin{figure*}[p]
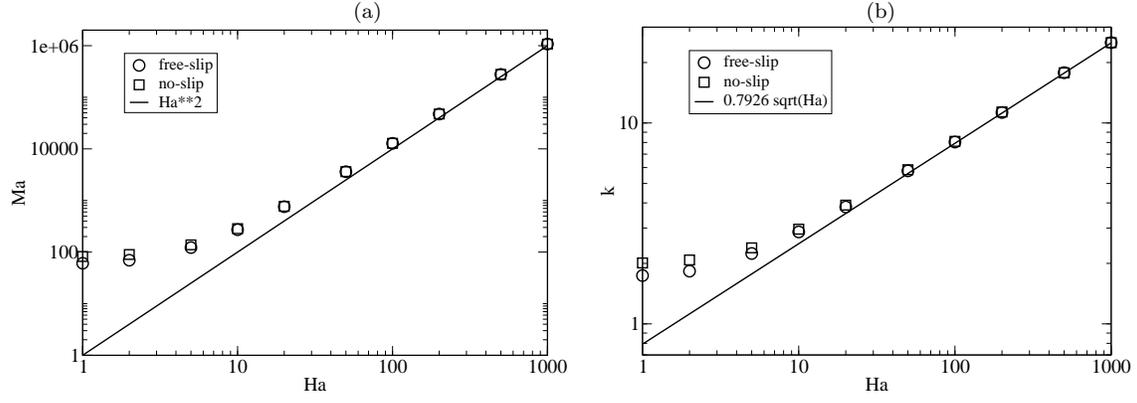

\scriptsize{
\parbox{0.5\linewidth}{(a)}\parbox{0.33\linewidth}{(b)}
\centering
\includegraphics[width=0.45\textwidth,clip=]{pics/mavsha.eps}
\includegraphics[width=0.45\textwidth,clip=]{pics/kvsha.eps}
\caption{\label{fig:linearstability}{Critical Marangoni and wavenumbers for 
neutral conditions. }}}
\end{figure*}

Critical conditions are found by searching for the minimum $Ma(k)$ for the
given value of $Ha$, which provides the critical parameters $Ma_c(Ha)$
and $k_c(Ha)$. The results are shown in Fig. \ref{fig:linearstability} for free-slip ($D^2W(0)=0$)  
and no-slip ($D W(0)=0$) boundary conditions at the bottom. The
critical Marangoni and wavenumber are smaller for the free-slip case, but 
the difference vanishes as $Ha$ increases. The asymptotic relations 
\begin{equation}
Ma_c\sim Ha^2,\hskip1cm k_c\sim 0.7926 \sqrt{Ha}, 
\end{equation}
shown in Fig. \ref{fig:linearstability} have been obtained by Wilson \cite{Wilson:1993} for the no-slip case
and are discussed and interpreted by Thess \& Nitschke 
\cite{Thess:Nitschke:1995}. They  appear
to be valid irrespective of the bottom boundary condition on the velocity.
The vertical structure of the neutral modes differs only in a zone of width
$1/Ha$ at the bottom between the free- and no-slip cases. 
Near the free surface, the unstable modes display
a boundary layer structure with a width of order $1/Ha$ as explained in 
\cite{Thess:Nitschke:1995}.

\section{Two-dimensional convection}

\subsection{Transition from weak to inertial convection}

\begin{figure*}[p]
\parbox{0.9\linewidth}{\hskip5mm (a)}
\centerline{
\includegraphics[width=0.33\textwidth,clip=]{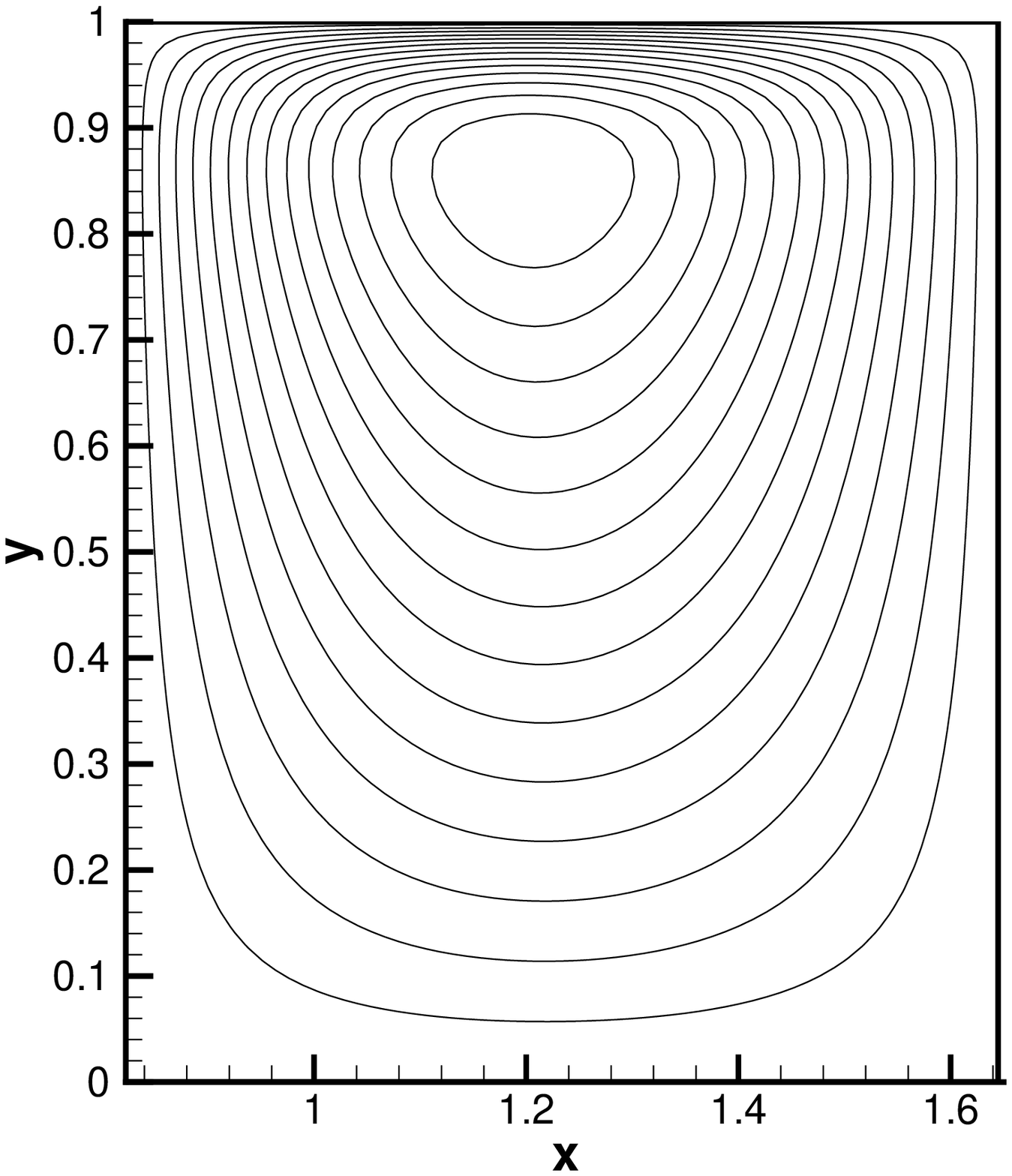}
\includegraphics[width=0.33\textwidth,clip=]{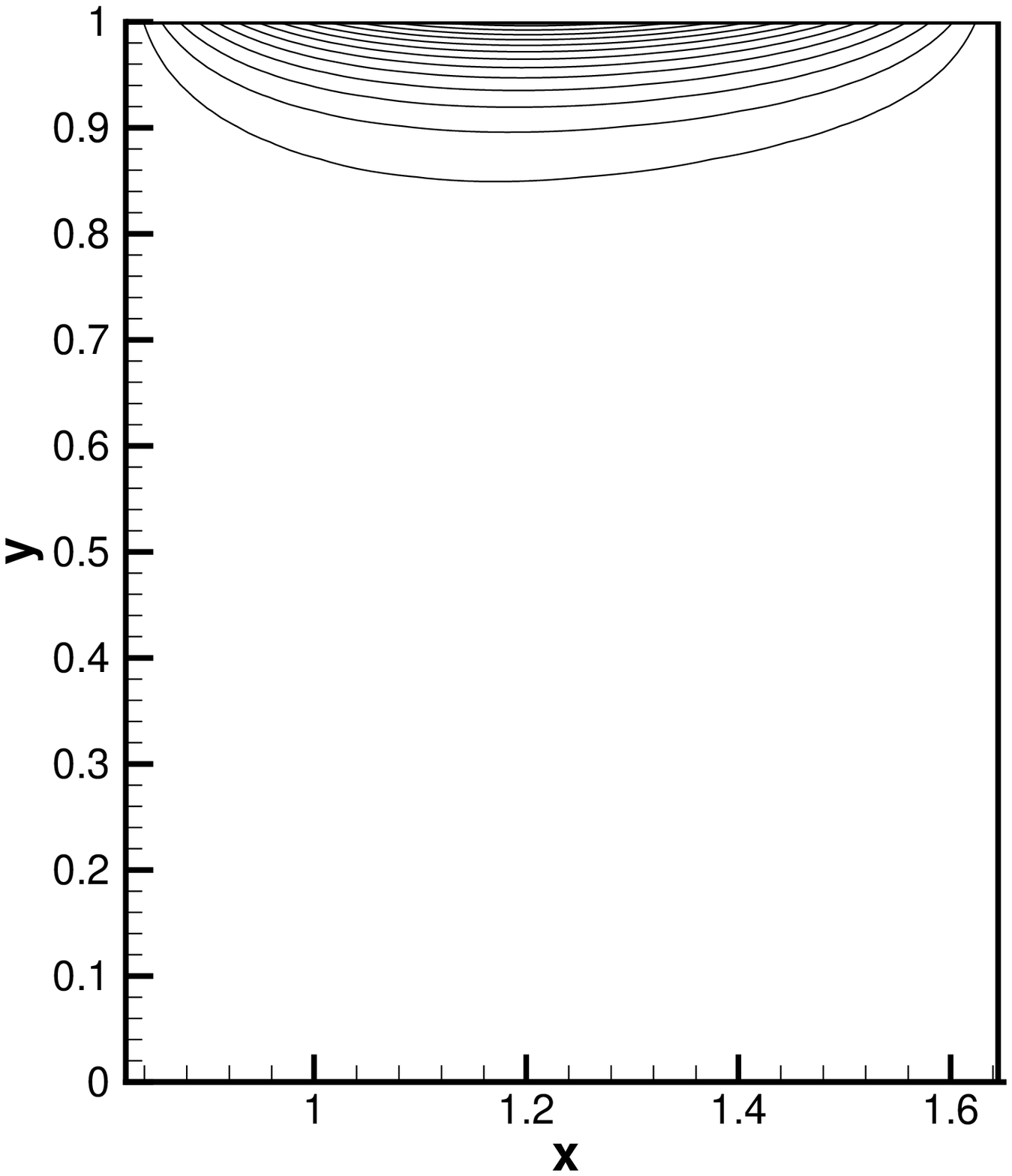}
\includegraphics[width=0.33\textwidth,clip=]{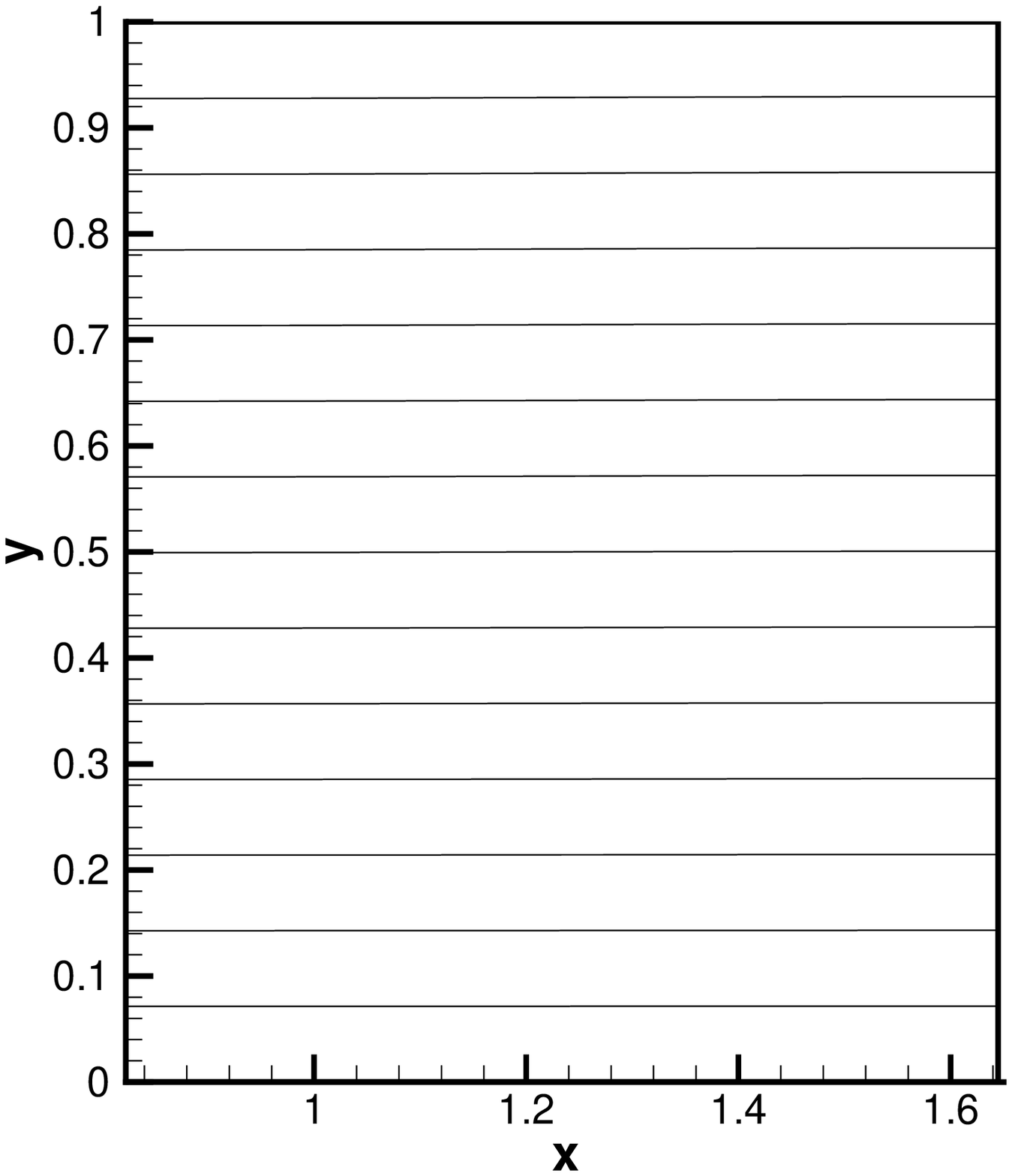}
}\hfill
\parbox{0.9\linewidth}{\hskip5mm (b)}
\centerline{
\includegraphics[width=0.33\textwidth,clip=]{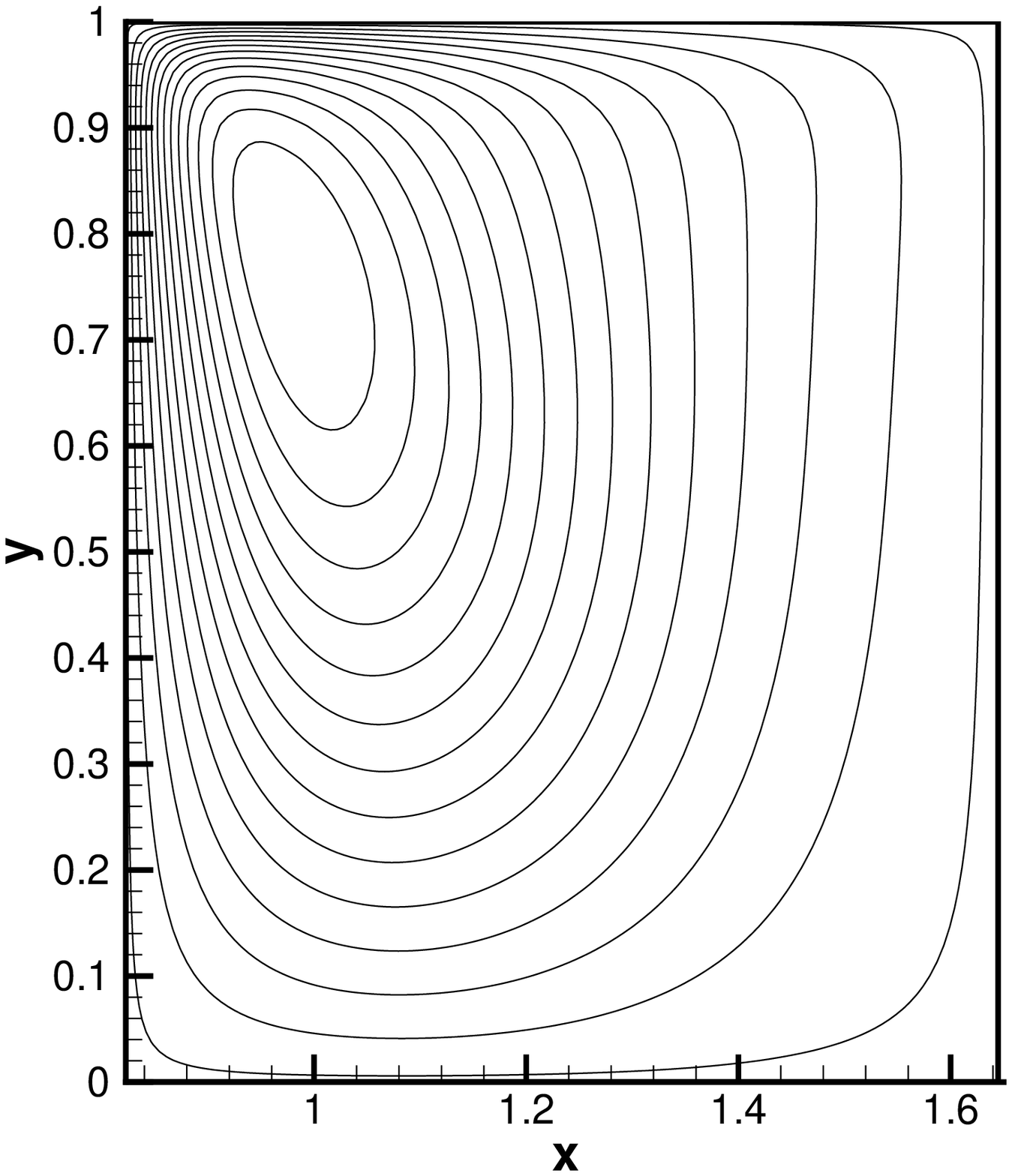}
\includegraphics[width=0.33\textwidth,clip=]{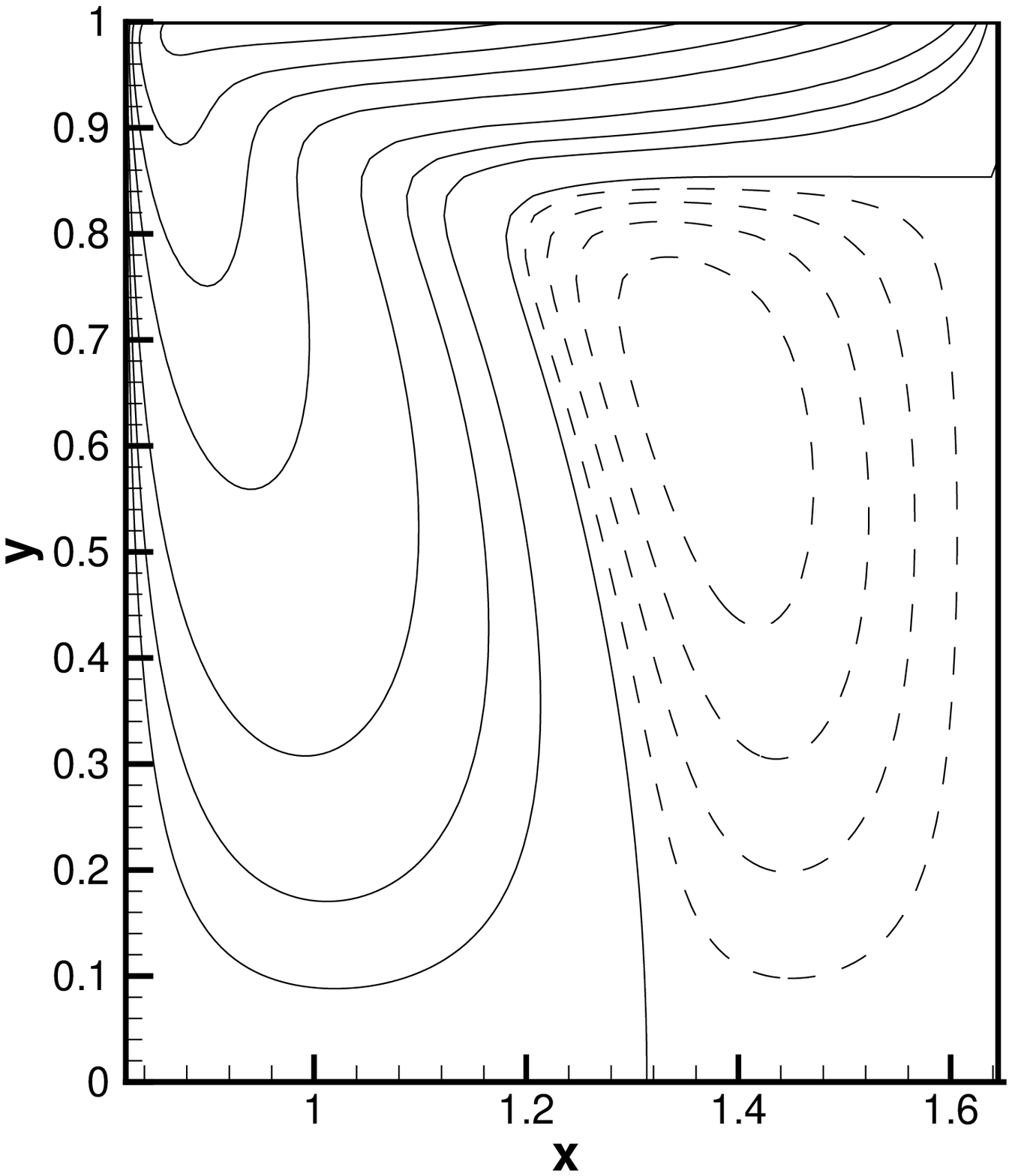}
\includegraphics[width=0.33\textwidth,clip=]{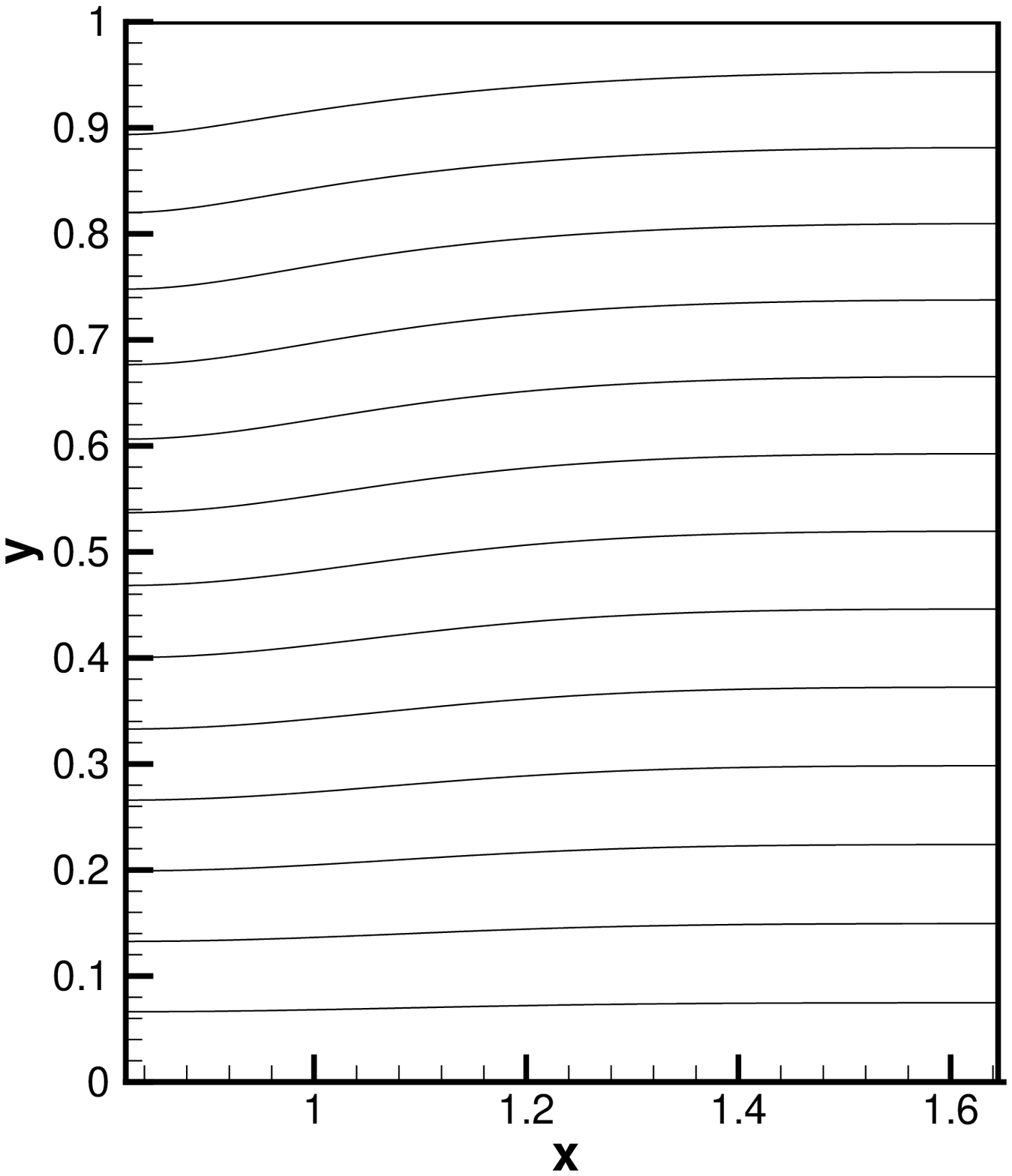}
}\hfill
\parbox{0.9\linewidth}{\hskip5mm (c)}
\centerline{
\includegraphics[width=0.33\textwidth,clip=]{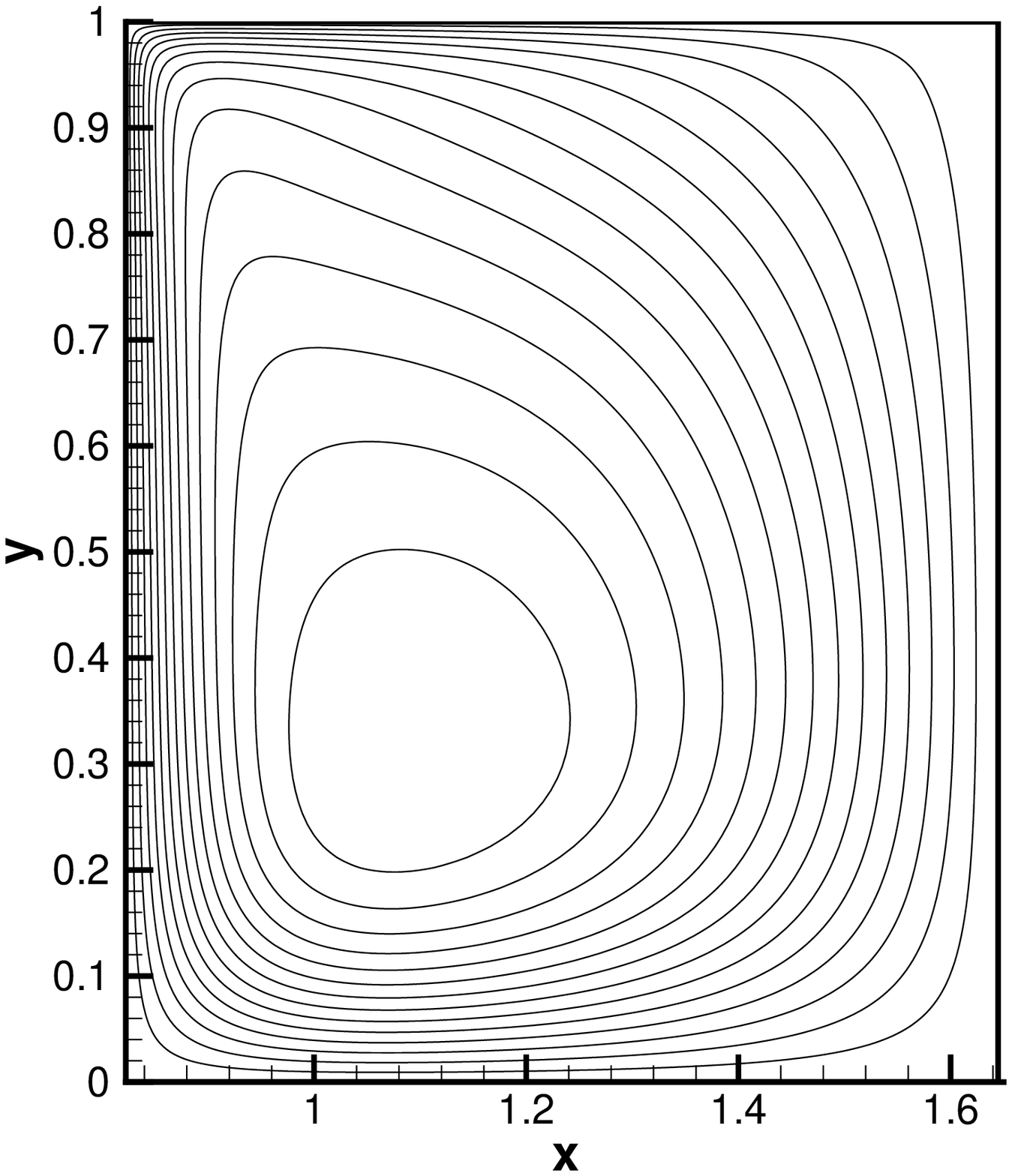}
\includegraphics[width=0.33\textwidth,clip=]{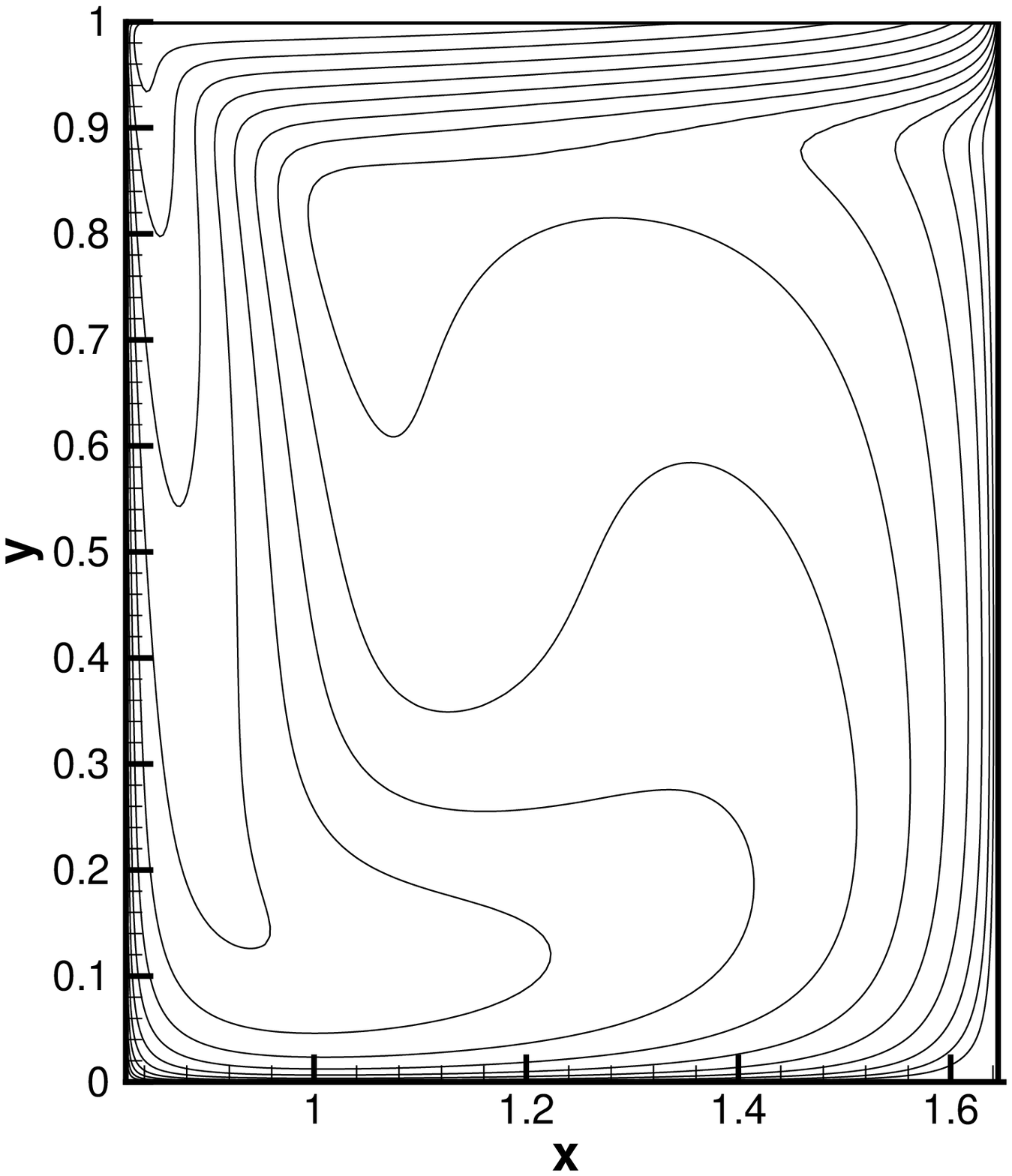}
\includegraphics[width=0.33\textwidth,clip=]{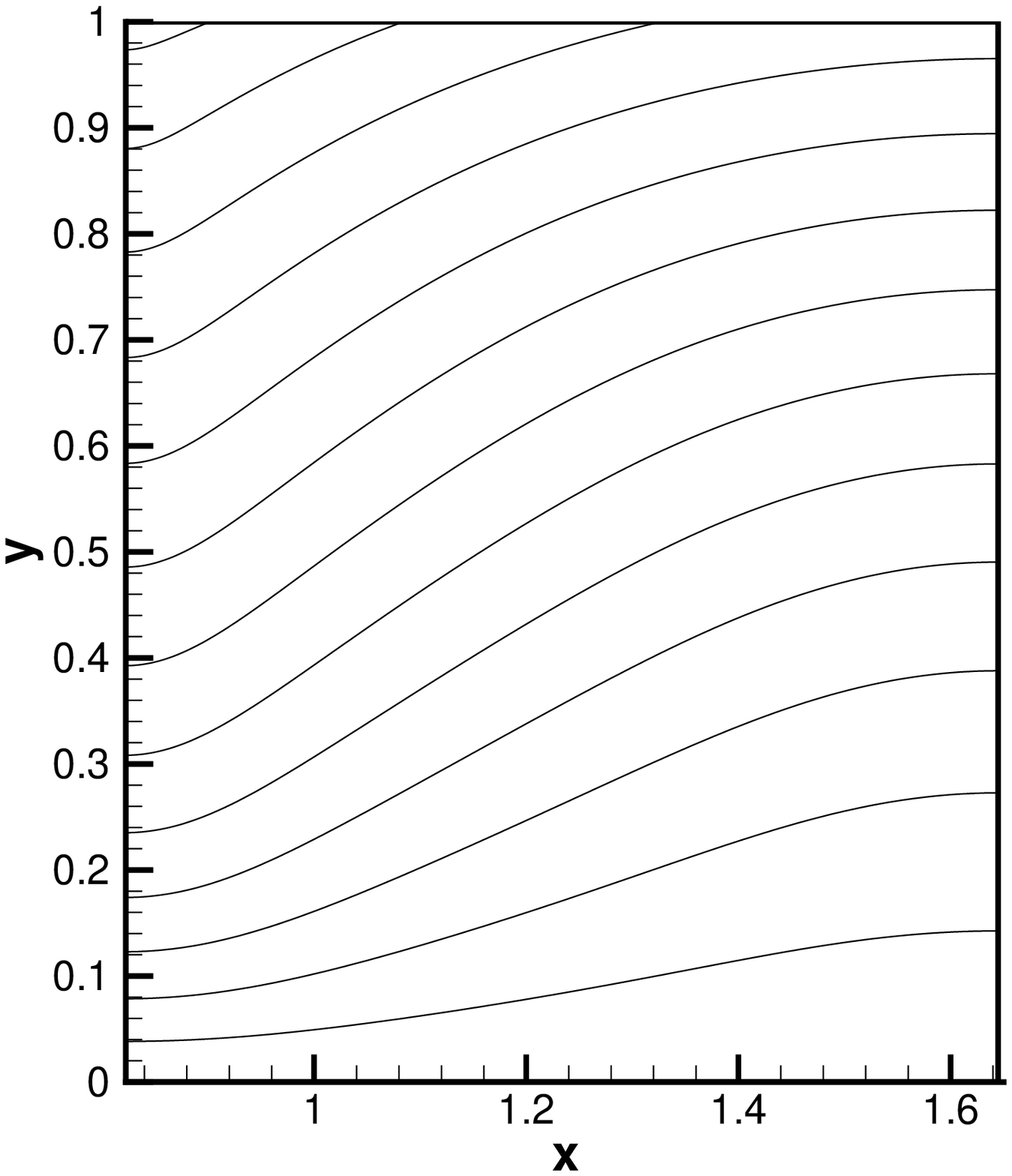}
}\hfill
\parbox{0.9\linewidth}{\hskip5mm (d)}
\centerline{
\includegraphics[width=0.33\textwidth,clip=]{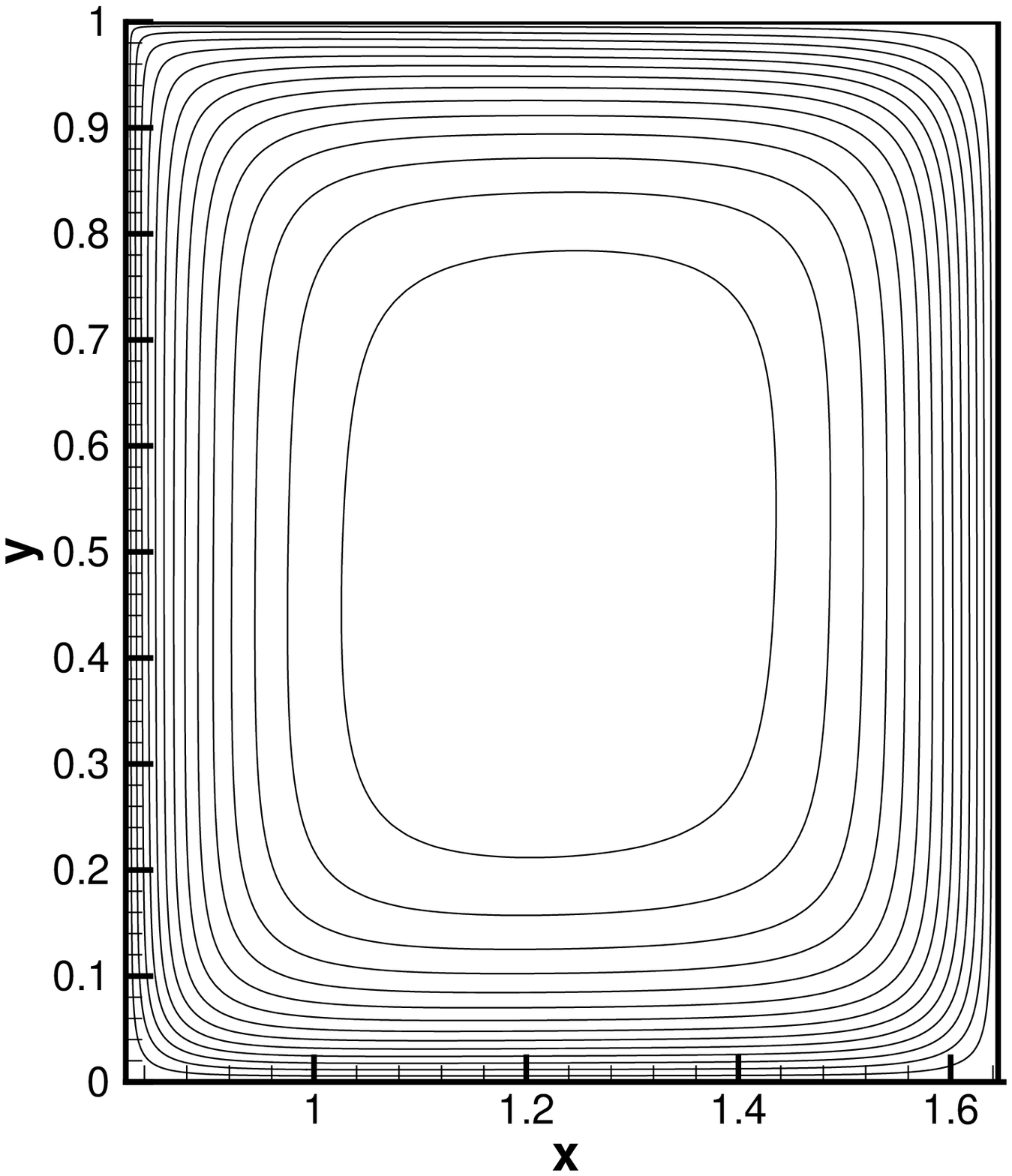}
\includegraphics[width=0.33\textwidth,clip=]{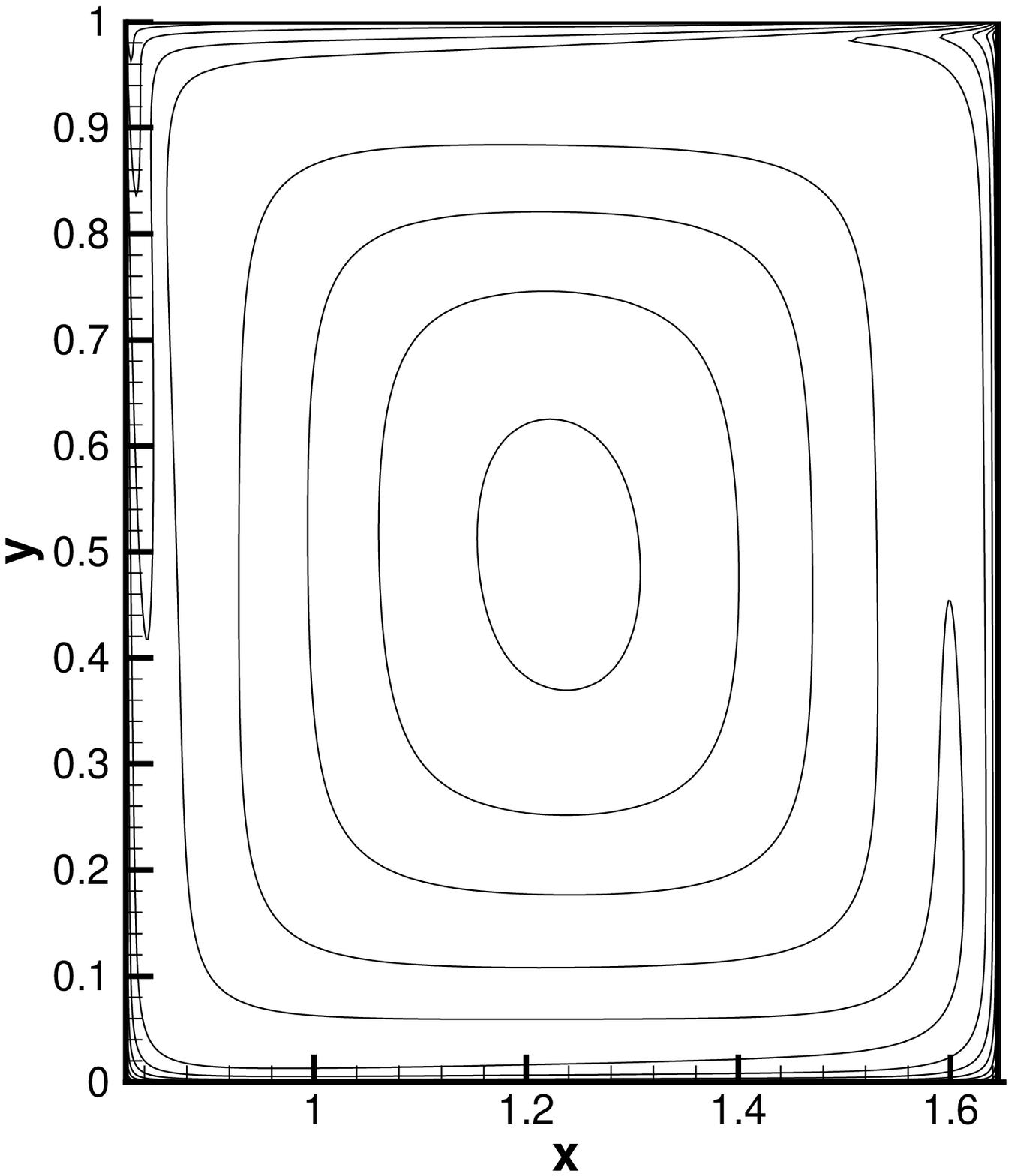}
\includegraphics[width=0.33\textwidth,clip=]{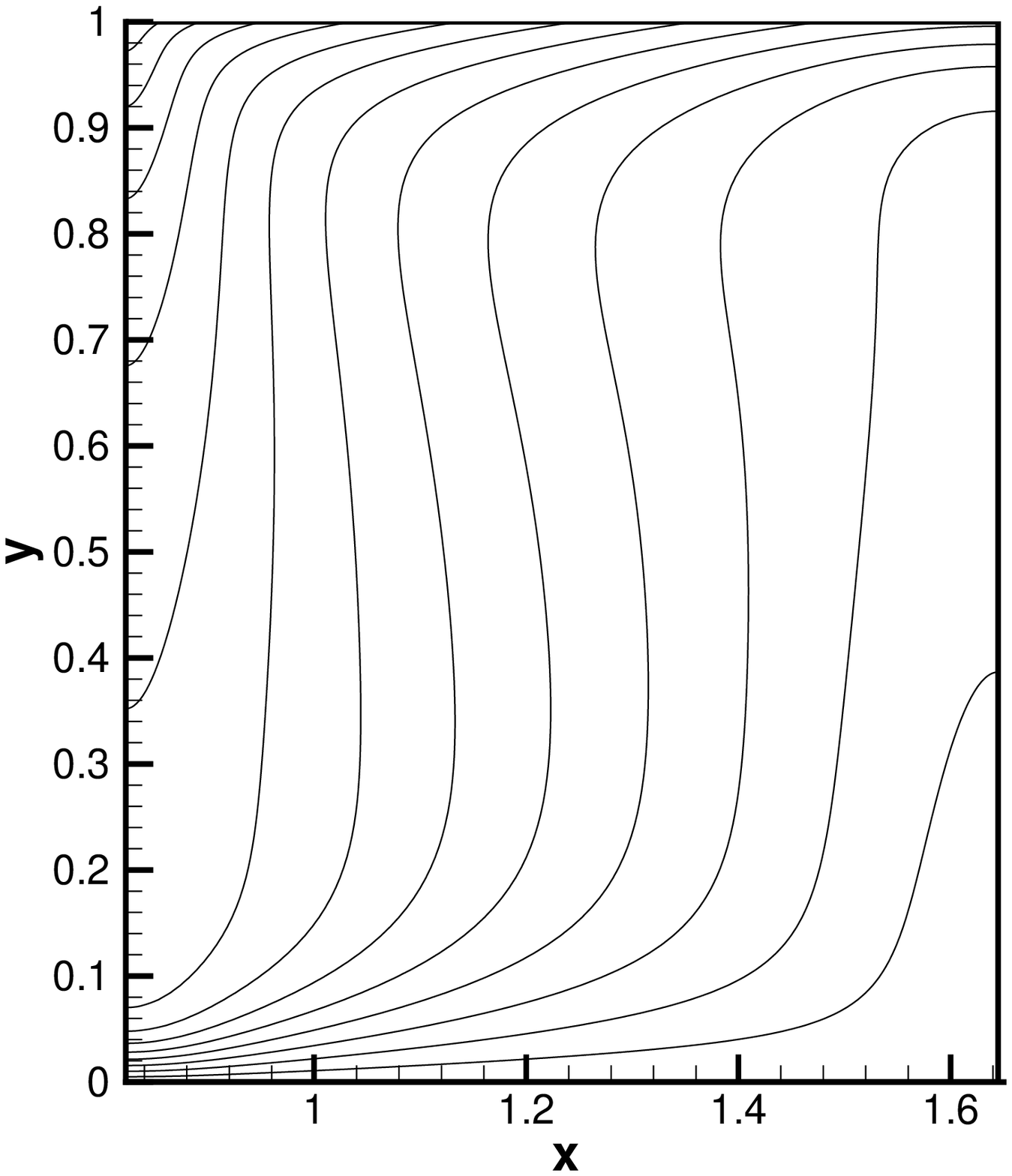}
}\hfill
\centerline{} 
\vskip-13mm
\caption{\label{fig:snapshots-ha20-pr02}{Isocontours of streamfunction (left column),
vorticity (middle column) and temperature (right column) of steady
two-dimensional convection with $Ha=20$,
$P=0.02$, $L_x=2\pi/k_c$. Only a single roll is shown due to the imposed mirror symmetry. The flow orientation
is counter-clockwise. (a) corresponds to $Ma=758$, (b) to $Ma=1000$, (c) to $Ma=2000$, (d) to $Ma=32000$. }}
\end{figure*}

Two-dimensional solutions will be considered first as they display several
interesting features in the non-magnetic case \cite{Boeck:1997}. 
In particular, large
Marangoni numbers can be explored more easily than in three-dimensional simulations.
The computational domains are chosen on the basis of the linear
stability analysis in the previous section for Hartmann numbers $Ha=10$ and $Ha=20$.
For these values we have the critical parameters $Ma_c(Ha=10)=271.2$, $k_c(Ha=10)=2.87$
and  $Ma_c(Ha=20)=757.2$, $k_c(Ha=20)=3.82$. The simulations are therefore performed with a
periodicity length $L_x=2\pi/k_c$, where $k_c$ takes the  value corresponding 
to $Ha=10$ and
$Ha=20$. This way, two roll cells fit into the computational domain at the onset of
convection, i.e. for $Ma_c$. In order to delay instabilities and to
maintain this flow topology at larger values of $Ma$
the mirror symmetry between the rolls is enforced by the numerical method through
symmetry conditions on the Fourier coefficients. 

Simulation results  are shown in Fig. \ref{fig:snapshots-ha20-pr02} 
for $Ha=20$ and $Pr=0.02$ for several values of the Marangoni number. 
The simulations evolve to a steady flow in all cases shown. Near the onset of
convection at $Ma=758$ nonlinear effects are weak, and the flow field reflects the structure
of the neutrally stable mode associated with the critical Marangoni number. 
One can clearly
identify the boundary layer at the free surface originating from the Lorentz force in the
streamline pattern.

For  $Ma=1000$ (Fig. \ref{fig:snapshots-ha20-pr02}(b)) 
the vorticity remains no longer confined near the free surface, but is
swept into the bulk because of the stronger advection. A significant difference to the
non-magnetic case is the change of sign of the vorticity in the domain indicated by
the dashed isolines (middle column). This change of sign  is an indication of the Lorentz
force impeding the advection of vorticity in the bulk of the liquid layer.
For larger values of $Ma$ the advection dominates and the negative vorticity region 
disappears. Moreover, the streamfunction  and  vorticity distributions become
approximately symmetric with  matching isolines near the center of the roll
(Fig. \ref{fig:snapshots-ha20-pr02}(d)). In this case, the flow approaches an inviscid
balance  between pressure gradient and advection term in the bulk region 
since the gradients of streamfunction and vorticity become aligned. The viscous and Lorentz
force terms are then weak by comparison with the other terms in the Navier-Stokes equation.

\begin{figure*}[ht]
\scriptsize{
\centering
\includegraphics[width=0.5\textwidth,clip=]{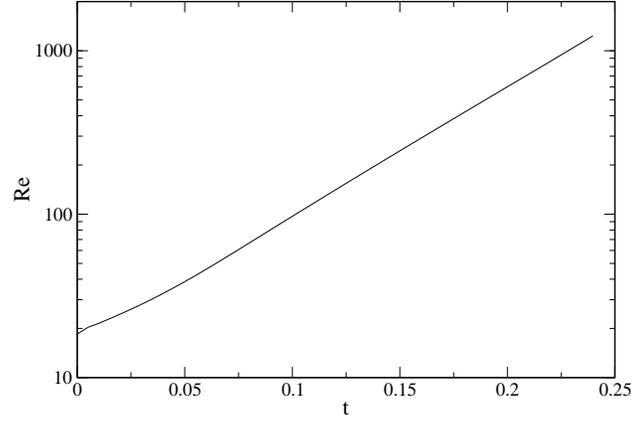}
\caption{\label{fig:reynoldsvstime}{Spatial rms velocity in viscous units
as function of time for two-dimensional BMC with $P=0$, $Ma=1350$ and $Ha=20$.  }}}
\end{figure*}

The tendency of the flow to approach such a nearly inviscid configuration has a profound
consequence for the limit of vanishing Prandtl number. When the forcing of the
flow by the surface tension gradients is strong enough to establish such a
nearly inviscid balance, then the approximate cancellation of pressure and advection term
imply that only the linear terms control the dynamics
in the momentum equation (\ref{nsviscous}).
The forcing is provided by the Marangoni boundary condition
coupling temperature perturbation and shear stress,
and the  temperature perturbation
becomes a linear functional of the velocity field in the limit $P=0$ (eq. (\ref{heatviscous})). 
One can then expect the velocity
to grow exponentially in time since the dynamics is governed by an effectively 
linear equation with a forcing term proportional to the velocity itself.
 This is illustrated by
Fig. \ref{fig:reynoldsvstime}, which shows the growth of the mean velocity with time for
$Ha=20$ and $Ma=1350$. The simulation was started from the steady solution for $P=0.005$
with the Prandtl number set to zero at $t=0$. After a short initial
transient, the  streamfuction and vorticity fields approach the symmetric state with
an approximate inviscid balance similar to Fig. \ref{fig:snapshots-ha20-pr02}(d), 
and the spatial rms velocity grows exponentially with time.
This behavior is completely analogous to the
non-magnetic case \cite{Boeck:1997}.

For finite Prandtl numbers, the unbounded growth of the velocity eventually
stops since the temperature difference across the layer is reduced by the 
fast advection, and the shear stress diminishes because of the correspondingly
reduced temperature difference on the free surface. Because the
saturation of the velocity is provided by the heat transport it turns out that
the appropriate scaling of the velocity and temperature
are provided by thermal units $\kappa/d$ and $\Delta T_0$. 
If velocity and temperature perturbations
are measured in these units, then these quantities become independent of $P$ for
$P\to 0$, and equations (\ref{nsthermal}-\ref{heatthermal}) are appropriate.
This mode of convection is the inertial convection mentioned in the introduction. 
Conversely, when the forcing by the surface-tension gradient is not too strong, 
then the velocity field does not attain the
symmetric streamlines required for the approximate cancellation of pressure gradient
and advection term, and the  momentum transport is responsible for nonlinear saturation of
the convective instability.
In this case, the viscous units provide the appropriate scales for 
velocity and temperature perturbation in the limit $P\to 0$. This mode of convection
has been termed weak convection in Ref.\cite{Boeck:1997}.

\begin{figure*}[ht]
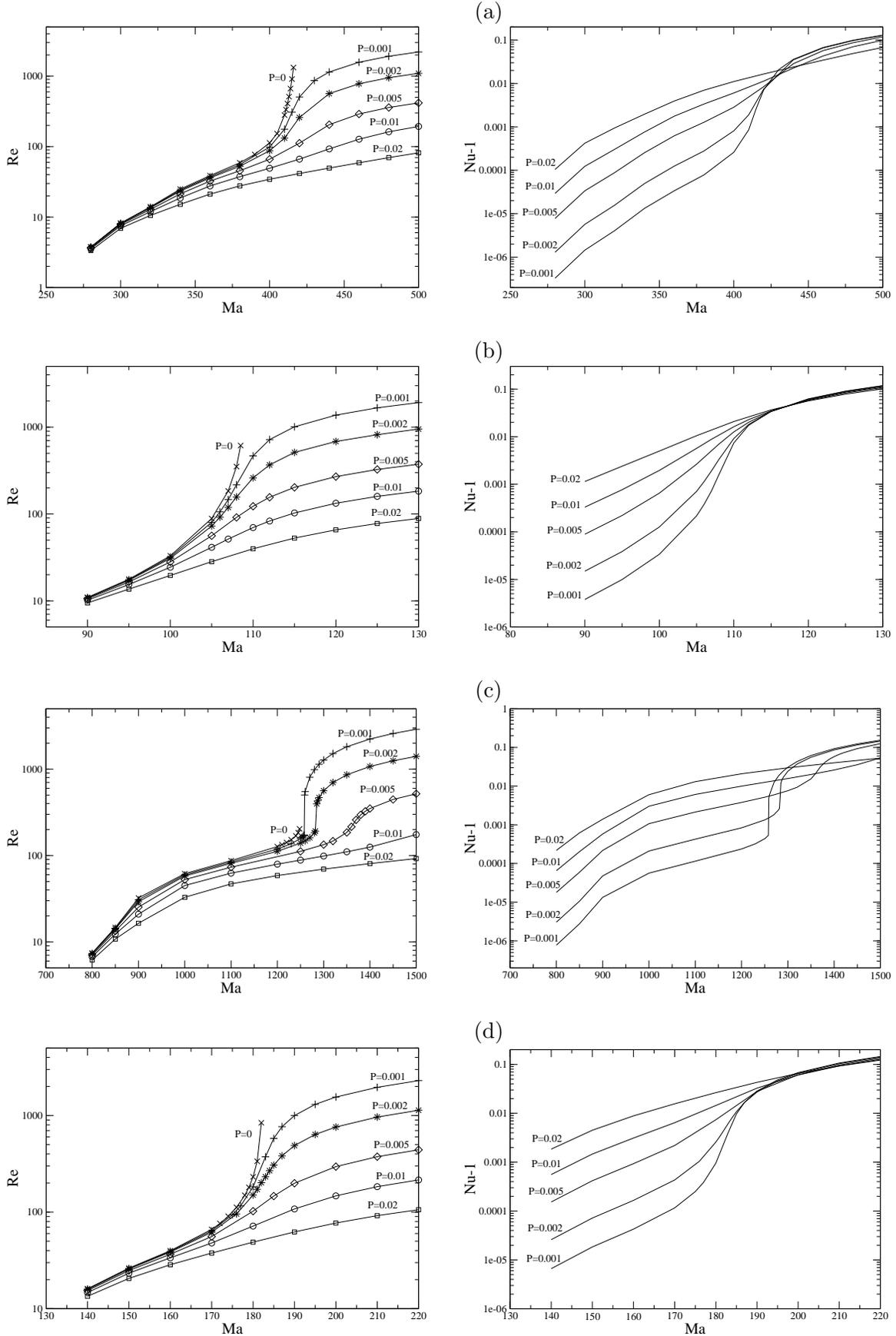

\parbox{0.9\linewidth}{\hskip15mm (a)}
\centerline{
\includegraphics[width=0.45\textwidth,clip=]{pics/RevsMa-Ha10-287.eps}
\hskip5mm\includegraphics[width=0.45\textwidth,clip=]{pics/Num1-Ha10-287.eps}
}\hfill
\parbox{0.9\linewidth}{\hskip15mm (b)}
\centerline{
\includegraphics[width=0.45\textwidth,clip=]{pics/RevsMa-Ha0-287.eps}
\hskip5mm\includegraphics[width=0.45\textwidth,clip=]{pics/Num1-Ha0-287.eps}
}\hfill
\parbox{0.9\linewidth}{\hskip15mm (c)}
\hfill\centerline{
\includegraphics[width=0.45\textwidth,clip=]{pics/RevsMa-Ha20-382.eps}
\hskip5mm\includegraphics[width=0.45\textwidth,clip=]{pics/Num1-Ha20-382.eps}
}\hfill
\parbox{0.9\linewidth}{\hskip15mm (d)}
\hfill\centerline{
\includegraphics[width=0.45\textwidth,clip=]{pics/RevsMa-Ha0-382.eps}
\hskip5mm\includegraphics[width=0.45\textwidth,clip=]{pics/Num1-Ha0-382.eps}
}\hfill
\vskip-5mm\caption{\label{fig:weaktoinertial}{Reynolds and Nusselt numbers for 
 steady two-dimensional BMC with $L_x=2\pi/k$. Cases (a,b) have $k=2.87$ and different
 values $Ha=10$  (a) and $Ha=0$ (b). Cases (c,d) have $k=3.82$ and different
 values $Ha=20$  (c) and $Ha=0$ (d). }}
\end{figure*}

The change from viscous to thermal scaling  appears in a certain range of
Marangoni numbers. Fig. \ref{fig:weaktoinertial} illustrates the switch from viscous to thermal scaling
by the Reynolds number, i.e. the spatial rms velocity in viscous units,
and the quantity $Nu-1$, which scales as $P^2$ when
the viscous scaling applies, and which becomes independent of $P$ when the thermal scaling
holds. The steady flows were computed for five finite values of the Prandtl number from
$P=0.001$ up to $P=0.02$ and for $P=0$, and for $Ha=10$ and $Ha=20$ with the
corresponding domain sizes $L_x=2\pi/k_c$. In addition, the computations were repeated for
the non-magnetic case ($Ha=0$) with
the same domain sizes $L_x=2\pi/k_c$ as for $Ha=10$ and $Ha=20$. 
The numerical resolution for the simulations was adjusted depending on the Reynolds number,
and convergence was verified for several parameter sets by doubling
the number of collocation points in both directions.
Typical resolutions at the lowest value of $P$ were $N_x=1024$ and $N_z=129$ collocation points.

The left column of Fig.\ref{fig:weaktoinertial} shows that
the spatial rms velocity (Reynolds number)
approaches the limiting curve for $P=0$ when the Marangoni number is not too large.
When scaled in thermal units, the rms velocity would tend to zero as $P\to 0$, i.e. the
curves would look similar to $Nu-1$ in the right column. For sufficiently large Marangoni
numbers, the curves for $Nu-1$ show a tendency to converge onto a limiting curve as $P$ is
reduced. The rms velocity of the steady solution branch for $P=0$ 
shows either a very rapid growth over a narrow $Ma$-range or appears to end abruptly for $Ha=20$.
In the non-magnetic case 
there is an additional bifurcation in
the $P=0$ solution branch \cite{Boeck:1997},
 which appears to be absent in the four combinations of $Ha$ and
$L_x$ investigated here. This bifurcation allows one to clearly distinguish between the
weak convection regime with viscous scaling for $P\to 0$  and the inertial  
regime with thermal scaling in the non-magnetic case \cite{Boeck:1997}. In the
present cases, the transition from weak to inertial convection  at fixed $P$ is gradual except for
$Ha=20$, where a discontinuity is present at sufficiently low $P$. Concerning the apparently 
different behaviors of the solution branches for $P=0$ one has to bear in mind that these
cases are very hard to resolve numerically. The solutions 
converge very slowly, and are rather sensitive to the spatial resolution, which
must be increased with the Reynolds number to resolve the increasingly smaller
viscous boundary layers. The 
asymptotic behavior with $Ma$ of the branches with $P=0$ cannot be reliably determined
in the present numerical approach.

The extensive computations summarized by Fig. \ref{fig:weaktoinertial} were
performed for the non-magnetic and magnetic cases with the same periodicity
length in order to quantify by how much the Lorentz  forces delays  
the transition from weak to inertial convection. 
To do so, the transition from weak to inertial
convection must be characterized by a typical Marangoni number. For the present
purposes, the characteristic Marangoni number $Ma_i$ will be identified by 
the crossing between the branches $Nu(Ma)$ for $P=0.001$ and $P=0.002$,
i.e.
\begin{equation}
Nu(Ma_i,P=0.001)=Nu(Ma_i,P=0.002)
\end{equation}
is used to describe the transition from weak to inertial convection. Table \ref{table1} lists the corresponding
values for the four cases presented in Fig.  \ref{fig:weaktoinertial}.
The quantity 
\begin{equation}
\epsilon_i=\frac{Ma_i -Ma_c}{Ma_c}\end{equation}
is larger for the magnetic than for the non-magnetic case, but the relative increase in $\epsilon_i$ 
between $Ha=10$ and $Ha=20$ is relatively moderate when compared with the increase in $\epsilon_i$ on
account of the changed domain size, i.e. comparing the two $\epsilon_i$ values for $Ha=0$.

\subsection{Behavior at large $Ma$}

\begin{figure*}[ht]
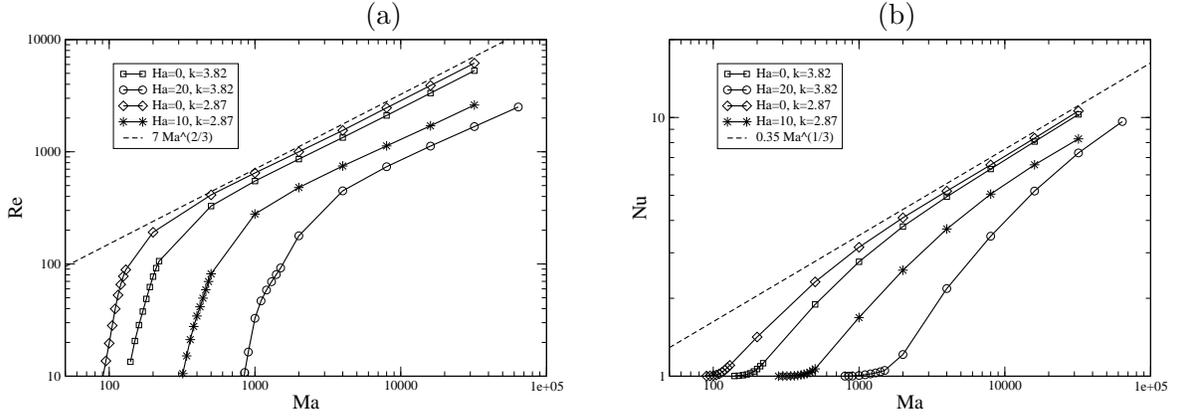

\parbox{0.5\linewidth}{(a)}\parbox{0.33\linewidth}{(b)}
\centering
\includegraphics[width=0.45\textwidth,clip=]{pics/RevsMa-P02-mod.eps}
\hskip5mm\includegraphics[width=0.45\textwidth,clip=]{pics/NuvsMa-P02-mod.eps}
\caption{\label{fig:powerlaws}{Reynolds and Nusselt numbers of steady two-dimensional BMC 
 for large $Ma$ with $P=0.02$ and $L_x=2\pi/k$.  }}
\end{figure*}

The behavior  of inertial  convection at  large values of $Ma$
is characterized by a boundary layer structure of temperature and vorticity
fields in the non-magnetic case, which leads to characteristic power laws 
\begin{equation}
\label{powerlaws}
V\sim Ma^{2/3}, \hskip5mm Nu\sim Ma^{1/3}
\end{equation}
for the
rms velocity $V$ (in thermal units)
and Nusselt number $Nu$. Since the Reynolds number is  $Re\sim V/P$,
it becomes very difficult to resolve the small structures in the flow
field for small $P$. The calculations for large $Ma$ have therefore only
been done for $P=0.02$ and for selected values of $Ma$ for $P=0.01$.
The results are shown in Fig. \ref{fig:powerlaws} for $P=0.02$. 
The two different domain sizes with $Ha=0$ show identical power law
behavior in agreement with eq. (\ref{powerlaws}), but have slightly different prefactors.

For non-zero $Ha$ there is no power-law scaling seen for $Nu$ in Fig.  \ref{fig:powerlaws}. 
For the rms velocity there is some indication, but with a smaller exponent than
for $Ha=0$. This may, however, be misleading because the temperature field
still lacks an isothermal core in the bulk, as can be seen in
 Fig. \ref{fig:snapshots-ha20-pr02}(d).
The absence of power-law scaling for $Nu$ is therefore not surprising.
Moreover, the estimates for the
kinetic energy dissipation given in Ref. \cite{Boeck:1997}
have to be corrected because of the additional
Lorentz force. It is not clear if a modified scaling would emerge from an attempt
to take the Lorentz force into account, which will not be made in this paper.

The Lorentz force has another interesting effect on the vorticity
distribution in inertial convection. In the non-magnetic case, one
finds an essentially constant vorticity inside the convection roll.
The physical argument for this constant value has been given by  Batchelor
\cite{Batchelor:1956}. Batchelor considers steady flow in
a region of closed streamlines with an
approximate balance of pressure gradients and advection terms in the momentum
equation, i.e. high-Reynolds-number flow. In this case, convection of vorticity
dominates along the streamlines, and the vorticity is therefore approximately
constant along  streamlines. Since the motion is assumed to be steady, viscous
diffusion of  vorticity across streamlines must vanish because 
the vorticity distribution would otherwise change with time.
The vorticity therefore cannot change between neighboring streamlines, and is
therefore constant in the entire region of high speed flow with closed
streamlines. 

\begin{figure*}[ht]
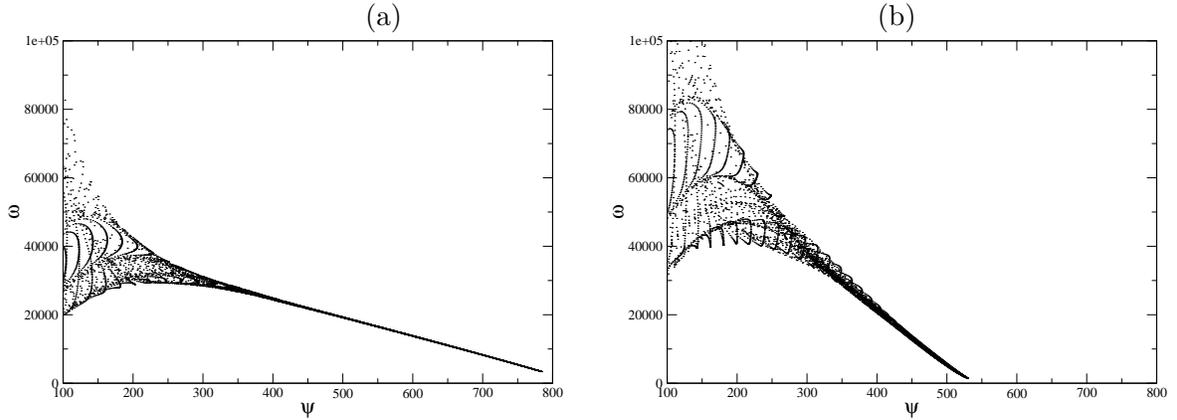

\parbox{0.5\linewidth}{(a)}\parbox{0.33\linewidth}{(b)}
\centering
\includegraphics[width=0.45\textwidth,clip=]{pics/psiomegaHa10-32000-002.eps}
\hskip5mm\includegraphics[width=0.45\textwidth,clip=]{pics/psiomegaHa20-64000-002.eps}
\caption{\label{fig:psiomega}{Vorticity-streamfunction scatter plots for 
steady two-dimensional BMC (in viscous units) with $Ha=10$, $Ma=32000$ (a)
and   $Ha=20$, $Ma=64000$ (b). The Prandtl number is $P=0.02$. }}
\end{figure*}

In magnetic inertial convection the additional Lorentz force is  small
when compared with pressure gradient and advection term, but it nevertheless
dissipates energy. In a steady state there must therefore be an influx
of energy into the region of closed streamlines. Since this influx must be provided 
by viscous diffusion, the vorticity cannot be constant as in the non-magnetic case. 
The scatter plots of vorticity and streamfunction shown in Fig. \ref{fig:psiomega}
demonstrate that this is indeed the case. The roll center is associated
with the highest values of $\psi$. Viscous effects dominate near the
roll boundaries where $\psi=0$, and no functional relation between
$\psi$ and $\omega$ exists for small values of $\psi$. Most of this region
has therefore been omitted from the scatter plots of 
Fig. \ref{fig:psiomega} by adjusting the $\psi$ axes.
For larger values of $\psi$ there is a clear functional relation between
vorticity and streamfunction, and  $\omega$ changes
almost linearly with $\psi$. The scatter is somewhat higher for $Ha=20$, which is
probably due to the lower Reynolds number caused by the stronger magnetic
damping.

The approximately linear behavior of $\omega$ with $\psi$ can be understood from a
modification of the derivation given by Batchelor \cite{Batchelor:1956}. The starting point
is the steady momentum (Navier-Stokes) equation in dimensional form,
\begin{equation}
\label{ns-dim}
\bm{v}\times\bm{\omega}=\nabla H
+\nu\nabla\times\bm{\omega}-\bm{f}/\rho,
\end{equation}
where the Lorentz force is denoted by $\bm{f}$ and
\begin{equation}
\label{totalhead}
H=p/\rho+\frac{q^2}{2}, \hskip5mm q^2=\bm{v}^2.
\end{equation}
The (approximate) inviscid balance implies that
\begin{equation}
\label{inviscidbalance}
\bm{v}\times\bm{\omega}=\nabla H.
\end{equation}
Following \cite{Batchelor:1956}, one now introduces orthogonal curvilinear
coordinates $(\psi,\xi)$, where the lines of constant $\xi$ are
orthogonal to the streamlines with  $\psi=const$. Differentials 
associated with $\psi$ and $\xi$ are $d\psi/q$ and $h d\xi$ in
physical space, where $h(\psi,\xi)$ is unknown.
The value of $H$ is constant along streamlines (Bernoulli equation), 
and the gradient of $H$
is therefore simply $q dH(\psi)/d\psi$. Equation (\ref{inviscidbalance})
then takes the form
\begin{equation}
\label{inviscidbalance2}
q\, \omega=q \, \frac{d H(\psi)}{d\psi},
\end{equation}
whereby the vorticity is a function of $\psi$ only. 

The next step is to integrate eq. (\ref{ns-dim}) around a closed
streamline.  The left hand side vanishes since 
the line element $d\bm{l}$ is everywhere parallel to $\bm{v}$ and therefore
orthogonal to $\bm{v}\times\bm{\omega}$. The first term on the right also
integrates to zero, i.e. one is left with 
\begin{equation}
\label{ns-dim2}
\nu \oint \nabla\times\bm{\omega} \cdot d\bm{l}=\frac{1}{\rho}\oint \bm{f} \cdot d\bm{l}
\end{equation}
The curl of $\omega$ is parallel to the streamline  because
$\omega(\psi)$, and the left hand side of eq. (\ref{ns-dim2})
simplifies to
\begin{equation}
\nu \oint \nabla\times\bm{\omega} \cdot d\bm{l}=\nu \,\frac{d\omega}{d\psi}\oint  
q dl.
\end{equation}
When the Lorentz force is zero, this equation implies that $\omega$ is
constant. 

For two-dimensional convection the Lorentz force is given by
\begin{equation}
\bm{f}=\bm{j}\times \bm{B}=\sigma\left(\bm{v}\times \bm{B}\right)\times \bm{B}=
\sigma B^2 \left[\bm{e}_z\left(\bm{e}_z\cdot\bm{v}\right)-\bm{v}\right]=
-\sigma B^2 v_x \bm{e}_x
\end{equation}
since the electric potential vanishes in this case. The integral on the right
hand side of eq. (\ref{ns-dim2}) then becomes
\begin{equation}
\label{lineintlorentz}
\oint \bm{f} \cdot d\bm{l}=-\sigma B^2  \oint v_x dx.
\end{equation}
The integral on the right hand side of eq. 
(\ref{lineintlorentz}) represents  a  fraction $C$ of the circulation
around the streamline.
One can therefore write eq. (\ref{ns-dim2}) as
\begin{equation}
\nu\, \frac{d\omega}{d\psi}\oint   q dl=
-\frac{\sigma B^2}{\rho} C \oint   q dl,
\end{equation}
i.e. the local slope of the vorticity-streamfunction dependence is
determined by  $\sigma B^2/\rho\nu$  and the prefactor $C$,
which depends  on the
streamline geometry and the velocity distribution along the streamline.
The almost linear decrease of $\omega$ with $\psi$ indicates that $C$ changes
only weakly with $\psi$.

\section{Three-dimensional convection}

As in the two-dimensional case, the focus of the present section 
is on the comparison with the non-magnetic case considered
earlier \cite{Boeck:1999}. 
The shape of the computational domain is therefore chosen
in agreement with this previous work, which focused on the smallest
rectangular domain compatible with a perfectly hexagonal
pattern at the onset of convection. It has the aspect ratios
$L_x=4\pi/k_c$, $L_y=4\pi/(\sqrt{3} k_c)$ with periodic 
boundary conditions in the $x$- and $y$-directions.
Because of the considerable computational expense of
three-dimensional simulations covering a significant
interval of Marangoni numbers the other parameters
are fixed to $P=0$ and $Ha=10$. The corresponding
wavenumber is $k_c=2.87$ as in the previous section.
As in the non-magnetic case, the limit $P=0$ is expected to 
provide a consistent
approximation for sufficiently small Prandtl numbers $P>0$.

\begin{figure*}[ht]
\parbox{0.5\linewidth}{(a)}\parbox{0.33\linewidth}{(b)}
\centering
\includegraphics[width=0.45\textwidth,clip=]{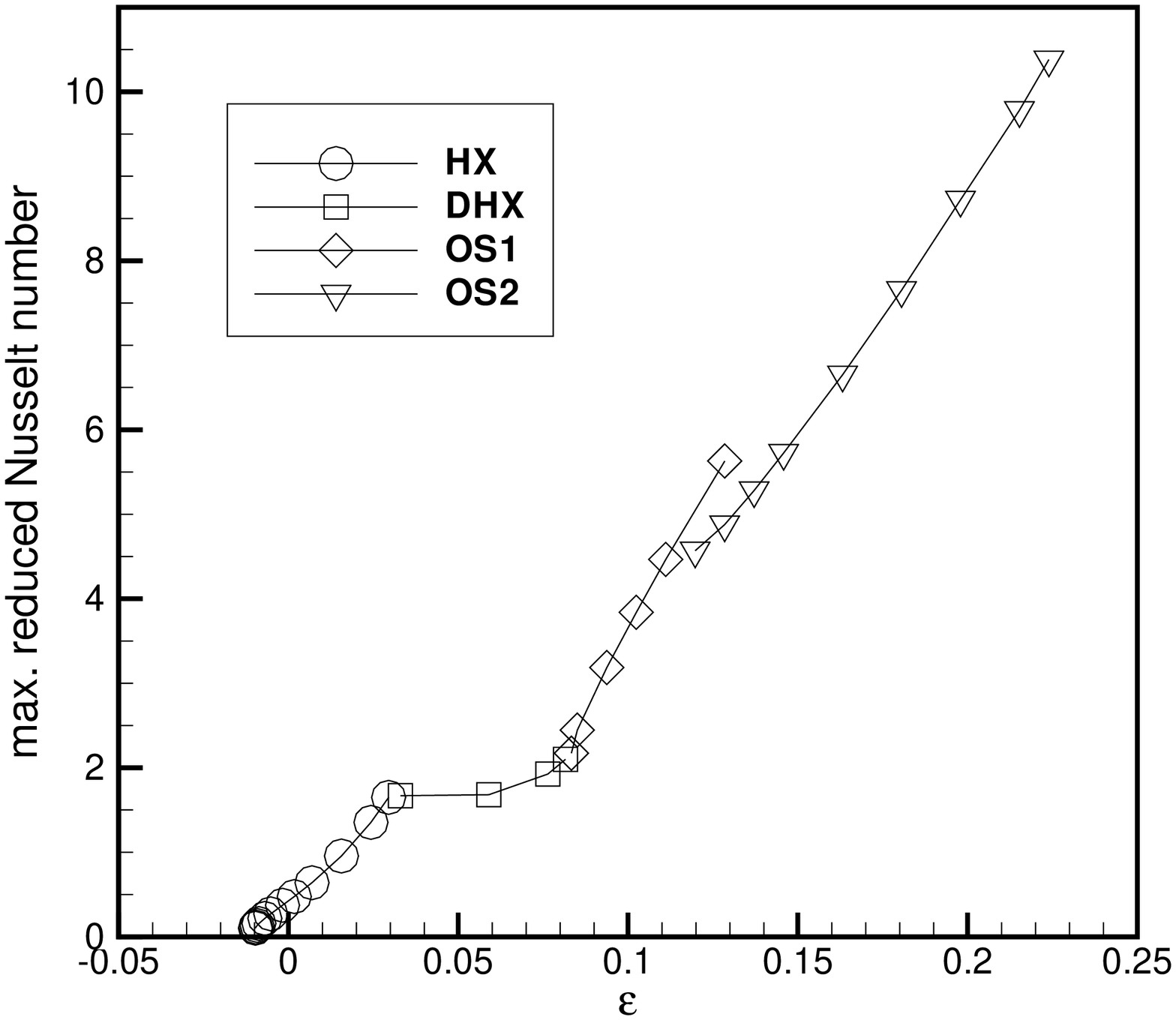}
\hskip5mm\includegraphics[width=0.45\textwidth,clip=]{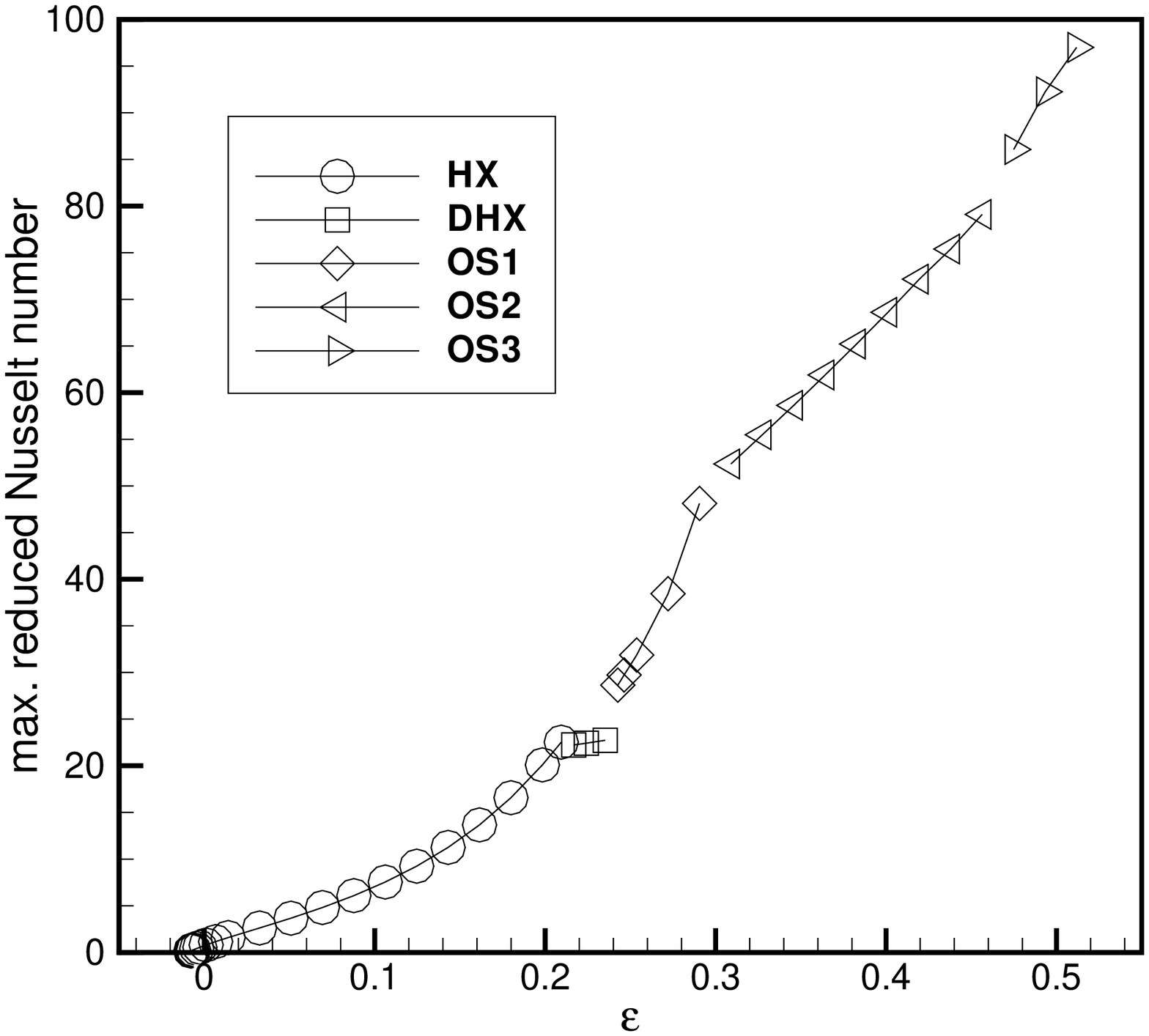}
\caption{\label{fig:3d-bif}{Solution branches for three-dimensional BMC
with $P=0$ in a rectangular cell with $L_x=4\pi/k$,$L_y=4\pi/\sqrt{3}k$. 
Parameters are $Ha=0$, $k=1.7$ (a) and $Ha=10$, $k=2.87$ (b). The different
branches are explained in the text.  Values of the reduced Nusselt number 
$\overline{v_z \theta}$ correspond to maxima  during the oscillation
period for time-dependent convection. }}
\end{figure*}

The simulations were performed by systematically increasing 
the Marangoni number using the final state of  previous 
runs as initial condition. Near $Ma_c$, the Marangoni
number was also decreased to explore the subcritical range.
Some simulations were started from
random initial conditions to check for additional 
solution branches, which may be missed by the 
continuation approach.
The main results of the simulations are summarized
in the plots of Fig. \ref{fig:3d-bif}, which show the 
reduced Nusselt number $\overline{v_z\theta}$
for different solutions 
as a function of the dimensionless parameter
\begin{equation}
\epsilon=\frac{Ma-Ma_c}{Ma_c}
\end{equation}
measuring the distance from the linear instability
threshold. The results for the
non-magnetic case from \cite{Boeck:1999} with
 $Ma_c=57.6$ and $k_c=1.70$ are shown
for comparison. 
 
Two different types of stationary solutions,
identified as perfect hexagons (HX) and deformed hexagons (DHX) are
found in both non-magnetic and magnetic cases. These solutions
are found up to $\epsilon\approx 0.24$ for $Ha=10$ whereas they only
exist up to $\epsilon\approx 0.08$ for $Ha=0$. The perfect hexagons
have hexagonal symmetry, which is lost in the deformed state.
As in the non-magnetic case, the 
hexagonal convection cells are characterized by 
downflow in the center of the hexagons. This flow orientation is different
from the hexagonal cells at $P>>1$, which have upflow in the
center of the hexagon \cite{Thess:Bestehorn:1995}.
The hexagons exist subcritically down to  $\epsilon\approx -0.008$
at $Ha=10$ and down to $\epsilon\approx -0.01$ at $Ha=0$.

Time-dependent solutions with oscillatory dynamics are marked by OS 
in Fig. \ref{fig:3d-bif}. As discussed in \cite{Boeck:1999},
the branch marked by OS1 corresponds to
an expansion/contraction of the cells in the $x$-direction. The solution branch 
OS2 displays an additional periodic excitation of the mean flow component
$V_y$ in the $y$-direction.  This solution branch slightly
overlaps with OS1. The OS2 branch leads to chaotic dynamics at
$\epsilon\approx 0.23$, which was explored in \cite{Boeck:2002}.

\begin{figure*}[ht]
\parbox{0.2\linewidth}{\hfill (a)}
\vskip2mm  
    \parbox{0.9\linewidth}{\centerline{
     \includegraphics[width=0.5\linewidth,clip=]{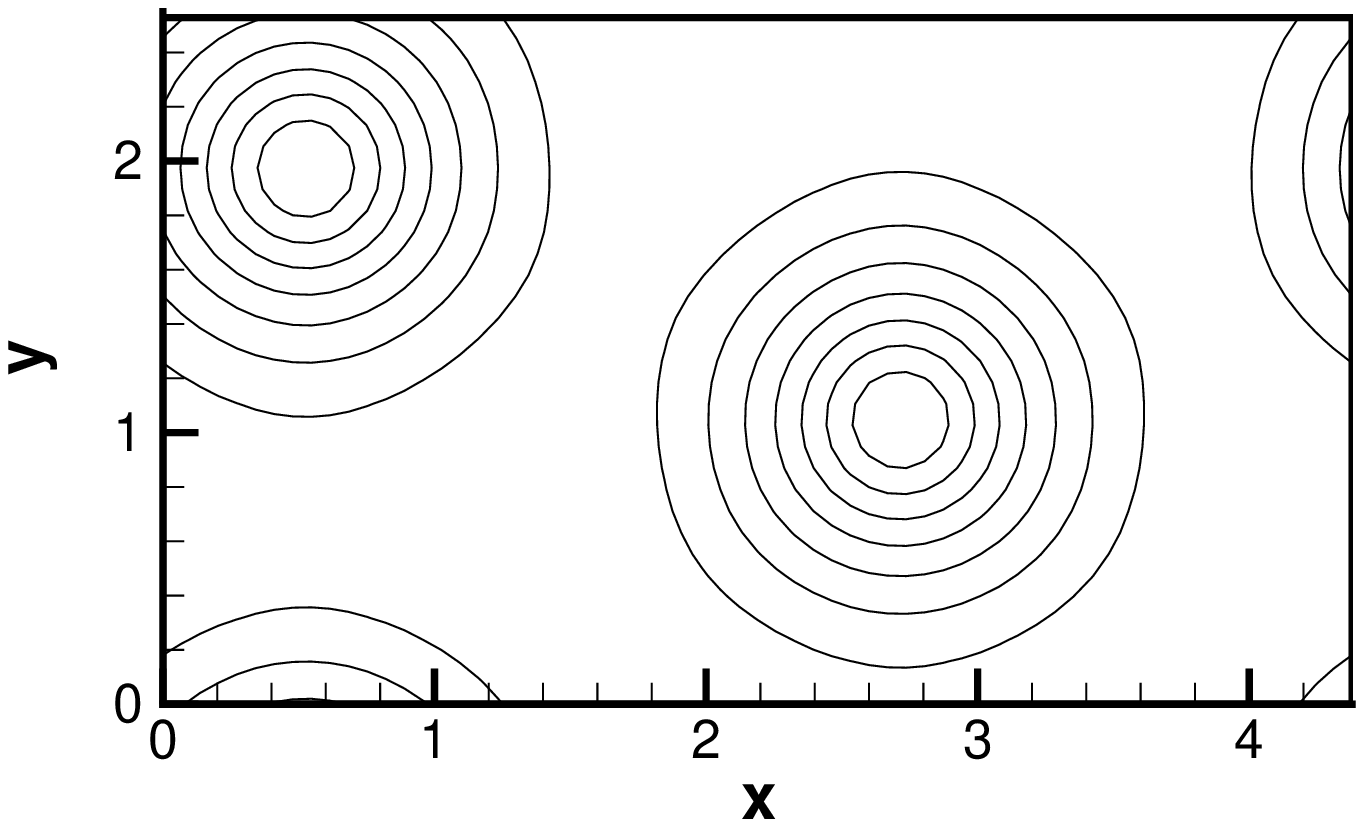}} 
    }
    
\vskip2mm
\parbox{0.2\linewidth}{ \hfill (b)}  
\vskip2mm
    \parbox{0.9\linewidth}{\centerline{
     \includegraphics[width=0.5\linewidth,clip=]{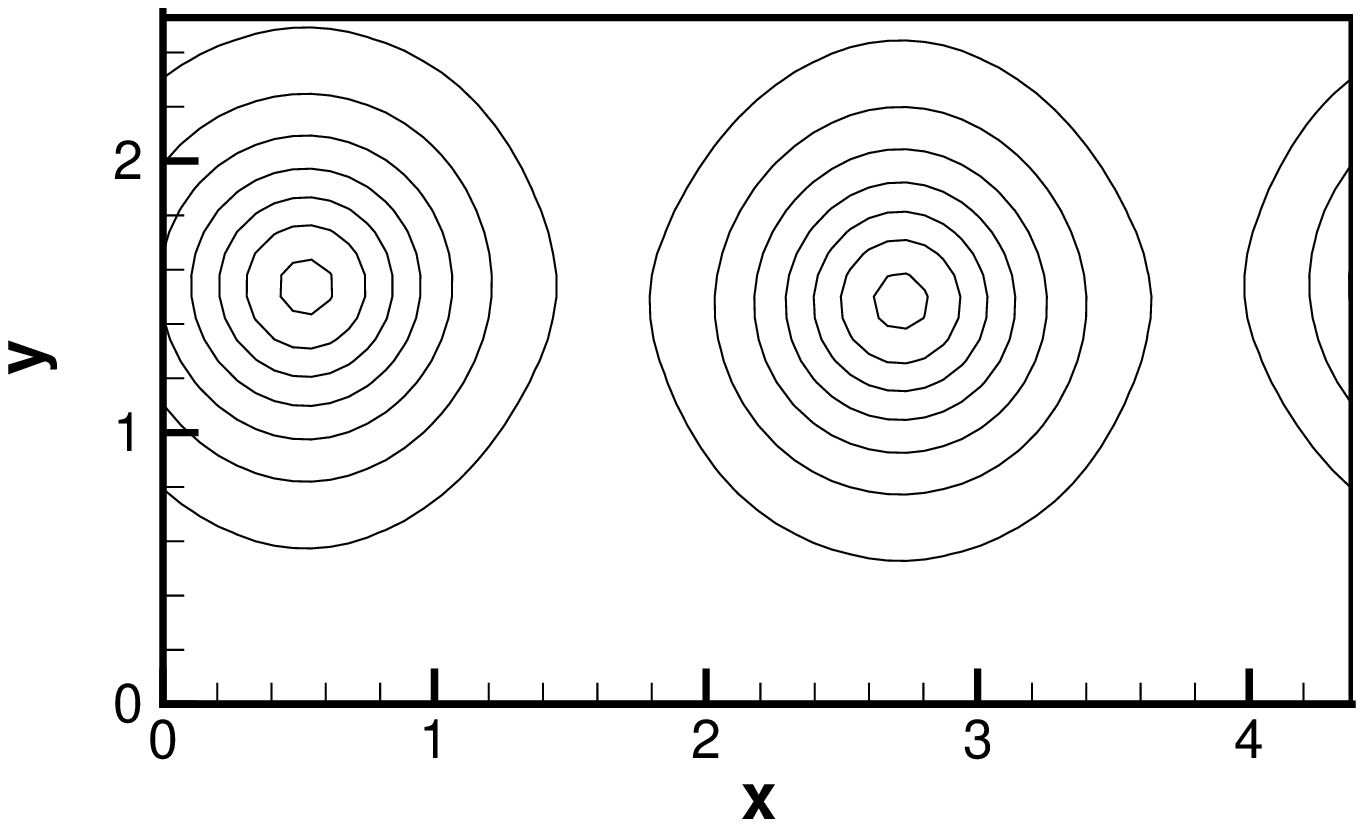}} 
    }
    \vskip2mm
\parbox{0.2\linewidth}{ \hfill (c)}  
\vskip2mm
    \parbox{0.9\linewidth}{\centerline{
     \includegraphics[width=0.5\linewidth,clip=]{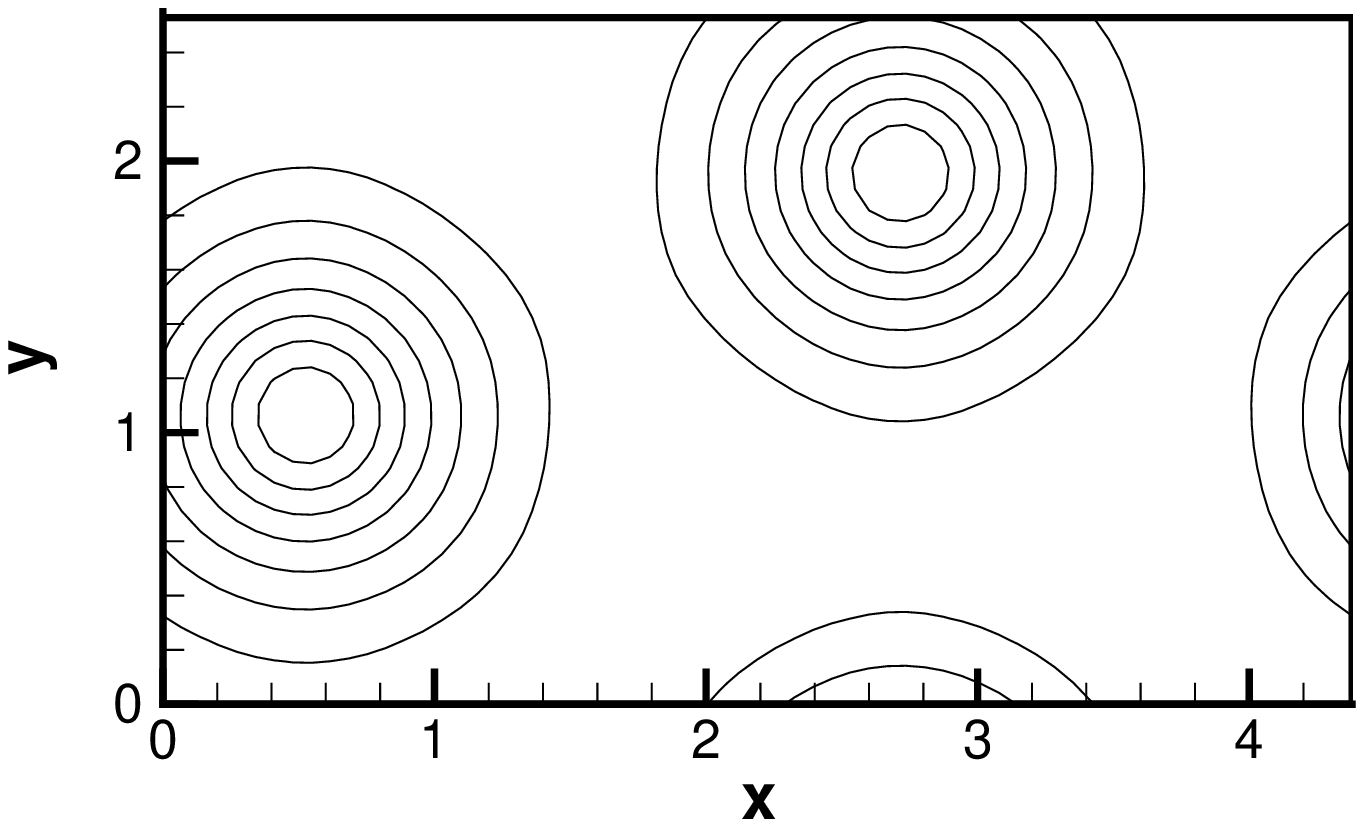}} 
    }
    \vskip2mm
\parbox{0.2\linewidth}{ \hfill (d)}  
\vskip2mm
    \parbox{0.9\linewidth}{\centerline{
     \includegraphics[width=0.5\linewidth,clip=]{pics/7d.eps}} 
    }
\caption{\label{fig:oscill-hex}Time-dependent BMC with $P=0$ and $Ha=10$: 
surface temperature perturbation  (a-c) and time evolution of
$\overline{v_z \theta}$ (d) for oscillating hexagons  at $Ma=345$.
The snapshots (a-c) correspond to the beginning, middle, and end of
the oscillation period shown in (d).
 }

\end{figure*}

\begin{figure*}[ht]
\parbox{0.2\linewidth}{\hfill (a)}
\vskip2mm  
    \parbox{0.9\linewidth}{\centerline{
     \includegraphics[width=0.5\linewidth,clip=]{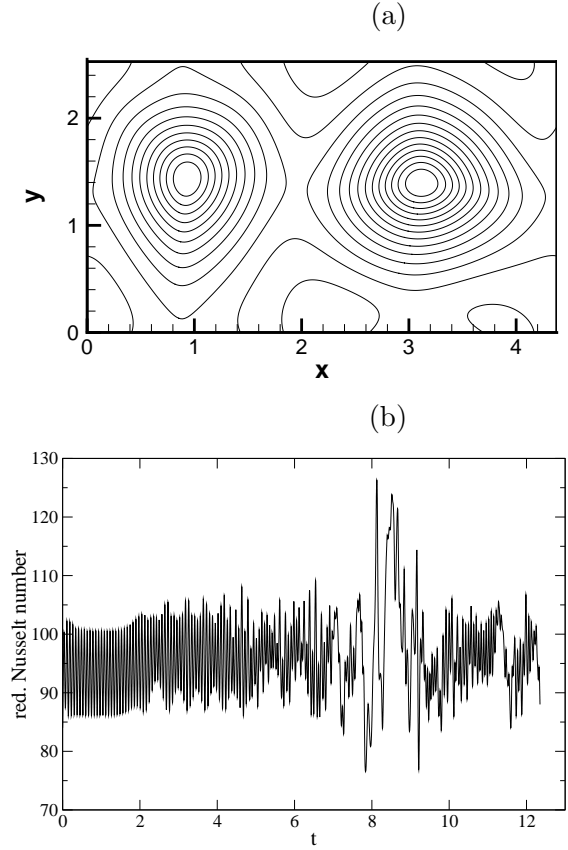}} 
    }
    
\parbox{0.2\linewidth}{\hfill (b)}    
\vskip2mm
    \parbox{0.9\linewidth}{\centerline{
     \includegraphics[width=0.5\linewidth,clip=]{pics/vztheta414.eps}} 
    }
\caption{\label{fig:chaos}Chaos in BMC with $P=0$ and $Ha=10$: 
surface temperature perturbation (a)  and time evolution of
$\overline{v_z \theta}$ (b)   at $Ma=414$. The simulation was
started from the periodic solution at $Ma=410$. }

\end{figure*}

For $Ha=10$ there are three different oscillatory
branches with regular dynamics. On the branch marked OS1,
the hexagons oscillate in $y$-direction and in antiphase
about their position as shown in Fig. \ref{fig:oscill-hex}.
During the oscillation, the hexagons
exchange their  $y$-position relative to each other, i.e.
the oscillation amplitude in the $y$ coordinate
is large even at the onset. The translatory motion  
in $y$  is accompanied by a size oscillation of the cells. 
On the OS2 branch,  the oscillatory motion in the $y$ 
direction changes its character. The hexagons do not
return to their initial position but continue their motion
in their respective direction. In effect, each cell
performs a size oscillation and travels in either the
positive or negative $y$-direction through the periodic
domain.
Finally, the OS3 branch is accompanied by an oscillation
of the $y$ component of the mean flow. On the 
OS2 branch the mean flow remains
zero. Chaotic dynamics appears beyond $\epsilon \approx 0.523$, i.e., $Ma\approx
413$. It is illustrated by Fig. \ref{fig:chaos}, which shows as snapshot of the surface
temperature perturbation, and the time series of the reduced Nusselt number.

In contrast to the non-magnetic case, no overlap could be found
for the different oscillatory branches, i.e. the simulations provided  
a unique oscillatory solution at given Marangoni number. The onset of time
dependent flow requires a higher supercriticality parameter $\epsilon$. The same
applies for the excitation of the mean flow and for chaos, i.e. the magnetic
field provides a considerable damping influence. Details of the transition to
chaos have not been explored so far due to the considerable cost of such
simulations. For $Ma\approx 400$ and larger ($\epsilon\approx 0.45$)
a numerical resultion of $N_x=128$, $N_y=64$ and
$N_z=33$ collocation points was needed in order to resolve  the 
structures of the velocity
field. The integration times were of the order $t_{\rm run}=10$
with a typical timestep $\Delta t=10^{-4}$.



\section{Conclusions}
The two- and three-dimensional simulations have systematically
explored the influence of a vertical magnetic field on 
low-Prandtl-number BMC. In addition to delaying the instability,
the magnetic field also extends the range of Marangoni numbers
where the zero-Prandtl-number limit works  for two-dimensional convection.
The transition from  weak to  inertial convection shows some
differences in comparison with the non-magnetic case of Ref.\cite{Boeck:1997},
but the phenomenon itself remains unchanged. Only the asymptotic power-law
scaling of inertial
convection with $Ma$ could not be detected at finite Hartmann numbers,
but may nonetheless be present when $Ma$ is increased further. The novel
feature of inertial convection with magnetic field is the modified 
vorticity distribution due to the magnetic damping.

The three-dimensional simulations were restricted to  $P=0$ and
$Ha=10$. The appearance of time-dependent flow and chaos is again delayed
by the magnetic damping, and the oscillatory solution branches differ from those
found at $Ha=0$. As already done for $Ha=0$, it could be interesting 
to examine the chaotization in more detail.

Throughout the paper only the free-slip boundary
condition at the bottom was considered. This was motivated by the
observation in Ref.\cite{Boeck:1997} that the free-slip
condition permits the existence of stationary 
rolls up to very large values of $Ma$, and thereby
allows one to study this asymptotic case rather
easily. No-slip conditions will lead to more complicated
behavior since  additional vorticity is produced at the
bottom wall. However, with the additional magnetic damping
the critical parameters for the free- and no-slip cases 
approach for large values of $Ha$, and the difference between
free- and no-slip could therefore be less significant than at $Ha=0$. 
The free-slip condition in the present paper was chosen mainly in order to
facilitate the comparison with the non-magnetic case. 

The work presented here suggests at least two directions for future
simulations. First, the no-slip condition and larger domain sizes
should be explored in order to see if the phenomena analyzed here remain
relevant in a  more realistic configuration. Second,  a horizontal
magnetic field should lead to interesting pattern dynamics due to the imposed
anisotropy. Ultimately, buoyancy and surface deformability should also be 
included in future numerical work.


\begin{table}[h]
\caption{Characteristic Marangoni numbers $Ma_i$ and nondimensional measures
$\epsilon_i$ for different parameter sets.}
\centering
\label{table1}       
\begin{tabular}{lllll}
\hline\noalign{\smallskip}
$Ha$ &  $k$ & $Ma_c$ & $Ma_i$ & $\epsilon_i$  \\[3pt]
\hline
$0$ & $2.87$ & $80.0$ & $115$ & $0.44$ \\
$10$ & $2.87$ & $271.2$ & $434$ & $0.60$ \\
$0$ & $3.82$ & $122.7$ & $187$ & $0.53$ \\
$20$ & $3.82$ & $757.2$ & $1259$ & $0.66$ \\
\noalign{\smallskip}\hline
\end{tabular}
\end{table}

\begin{acknowledgments}
The author acknowledges financial support
from the Deutsche Forschungsgemeinschaft in the framework of the
Emmy--Noether Program (grant Bo 1668/2). Computer resources were provided 
by  the computing centers of TU Ilmenau and TU Dresden.
\end{acknowledgments}



\begin{thebibliography}{30}
\expandafter\ifx\csname natexlab\endcsname\relax\def\natexlab#1{#1}\fi
\expandafter\ifx\csname bibnamefont\endcsname\relax
  \def\bibnamefont#1{#1}\fi
\expandafter\ifx\csname bibfnamefont\endcsname\relax
  \def\bibfnamefont#1{#1}\fi
\expandafter\ifx\csname citenamefont\endcsname\relax
  \def\citenamefont#1{#1}\fi
\expandafter\ifx\csname url\endcsname\relax
  \def\url#1{\texttt{#1}}\fi
\expandafter\ifx\csname urlprefix\endcsname\relax\def\urlprefix{URL }\fi
\providecommand{\bibinfo}[2]{#2}
\providecommand{\eprint}[2][]{\url{#2}}

\bibitem[{\citenamefont{Bestehorn}(1993)}]{bestehorn1993}
\bibinfo{author}{\bibfnamefont{M.}~\bibnamefont{Bestehorn}},
  \bibinfo{journal}{Phys. Rev. E} \textbf{\bibinfo{volume}{48}},
  \bibinfo{pages}{3622} (\bibinfo{year}{1993}).

\bibitem[{\citenamefont{Golovin et~al.}(1997)\citenamefont{Golovin,
  Nepomnyashchy, and Pismen}}]{Golovin:1997}
\bibinfo{author}{\bibfnamefont{A.~A.} \bibnamefont{Golovin}},
  \bibinfo{author}{\bibfnamefont{A.~A.} \bibnamefont{Nepomnyashchy}},
  \bibnamefont{and} \bibinfo{author}{\bibfnamefont{L.~M.}
  \bibnamefont{Pismen}}, \bibinfo{journal}{J. Fluid Mech.}
  \textbf{\bibinfo{volume}{341}}, \bibinfo{pages}{317} (\bibinfo{year}{1997}).

\bibitem[{\citenamefont{VanHook et~al.}(1997)\citenamefont{VanHook, Schatz,
  Swift, McCormick, and Swinney}}]{VanHook:1997}
\bibinfo{author}{\bibfnamefont{S.~J.} \bibnamefont{VanHook}},
  \bibinfo{author}{\bibfnamefont{M.~F.} \bibnamefont{Schatz}},
  \bibinfo{author}{\bibfnamefont{J.~B.} \bibnamefont{Swift}},
  \bibinfo{author}{\bibfnamefont{W.~D.} \bibnamefont{McCormick}},
  \bibnamefont{and} \bibinfo{author}{\bibfnamefont{H.~L.}
  \bibnamefont{Swinney}}, \bibinfo{journal}{J. Fluid Mech.}
  \textbf{\bibinfo{volume}{345}}, \bibinfo{pages}{45} (\bibinfo{year}{1997}).

\bibitem[{\citenamefont{Eckert et~al.}(1998)\citenamefont{Eckert, Bestehorn,
  and Thess}}]{Eckert:1998}
\bibinfo{author}{\bibfnamefont{K.}~\bibnamefont{Eckert}},
  \bibinfo{author}{\bibfnamefont{M.}~\bibnamefont{Bestehorn}},
  \bibnamefont{and} \bibinfo{author}{\bibfnamefont{A.}~\bibnamefont{Thess}},
  \bibinfo{journal}{J. Fluid Mech.} \textbf{\bibinfo{volume}{358}},
  \bibinfo{pages}{149} (\bibinfo{year}{1998}).

\bibitem[{\citenamefont{Davis}(1987)}]{Davis:1987}
\bibinfo{author}{\bibfnamefont{S.~H.} \bibnamefont{Davis}},
  \bibinfo{journal}{Annu. Rev. Fluid Mech.} \textbf{\bibinfo{volume}{19}},
  \bibinfo{pages}{403} (\bibinfo{year}{1987}).

\bibitem[{\citenamefont{Pumir and Blumenfeld}(1996)}]{Pumir:Blumenfeld:1996}
\bibinfo{author}{\bibfnamefont{A.}~\bibnamefont{Pumir}} \bibnamefont{and}
  \bibinfo{author}{\bibfnamefont{L.}~\bibnamefont{Blumenfeld}},
  \bibinfo{journal}{Phys. Rev. E} \textbf{\bibinfo{volume}{54}},
  \bibinfo{pages}{R4528} (\bibinfo{year}{1996}).

\bibitem[{\citenamefont{Kuhlmann and Rath}(1993)}]{Kuhlmann:Rath:1993}
\bibinfo{author}{\bibfnamefont{H.~C.} \bibnamefont{Kuhlmann}} \bibnamefont{and}
  \bibinfo{author}{\bibfnamefont{H.~J.} \bibnamefont{Rath}},
  \bibinfo{journal}{J. Fluid Mech.} \textbf{\bibinfo{volume}{247}},
  \bibinfo{pages}{247} (\bibinfo{year}{1993}).

\bibitem[{\citenamefont{Levenstam and Amberg}(1995)}]{Levenstam:Amberg:1995}
\bibinfo{author}{\bibfnamefont{M.}~\bibnamefont{Levenstam}} \bibnamefont{and}
  \bibinfo{author}{\bibfnamefont{G.}~\bibnamefont{Amberg}},
  \bibinfo{journal}{J. Fluid Mech.} \textbf{\bibinfo{volume}{297}},
  \bibinfo{pages}{357} (\bibinfo{year}{1995}).

\bibitem[{\citenamefont{Ginde et~al.}(1989)\citenamefont{Ginde, Gill, and
  Verhoeven}}]{Ginde:1989}
\bibinfo{author}{\bibfnamefont{R.~M.} \bibnamefont{Ginde}},
  \bibinfo{author}{\bibfnamefont{W.~N.} \bibnamefont{Gill}}, \bibnamefont{and}
  \bibinfo{author}{\bibfnamefont{J.~D.} \bibnamefont{Verhoeven}},
  \bibinfo{journal}{Chem. Eng. Comm.} \textbf{\bibinfo{volume}{82}},
  \bibinfo{pages}{223} (\bibinfo{year}{1989}).

\bibitem[{\citenamefont{Dauby et~al.}(1993)\citenamefont{Dauby, Lebon, Colinet,
  and Legros}}]{Dauby:1993}
\bibinfo{author}{\bibfnamefont{P.~C.} \bibnamefont{Dauby}},
  \bibinfo{author}{\bibfnamefont{G.}~\bibnamefont{Lebon}},
  \bibinfo{author}{\bibfnamefont{P.}~\bibnamefont{Colinet}}, \bibnamefont{and}
  \bibinfo{author}{\bibfnamefont{J.~C.} \bibnamefont{Legros}},
  \bibinfo{journal}{Q. Jl. Mech. appl. Math.} \textbf{\bibinfo{volume}{46}},
  \bibinfo{pages}{683} (\bibinfo{year}{1993}).

\bibitem[{\citenamefont{Thess and Bestehorn}(1995)}]{Thess:Bestehorn:1995}
\bibinfo{author}{\bibfnamefont{A.}~\bibnamefont{Thess}} \bibnamefont{and}
  \bibinfo{author}{\bibfnamefont{M.}~\bibnamefont{Bestehorn}},
  \bibinfo{journal}{Phys. Rev. E} \textbf{\bibinfo{volume}{52}},
  \bibinfo{pages}{6358} (\bibinfo{year}{1995}).

\bibitem[{\citenamefont{Boeck and Thess}(1997)}]{Boeck:1997}
\bibinfo{author}{\bibfnamefont{T.}~\bibnamefont{Boeck}} \bibnamefont{and}
  \bibinfo{author}{\bibfnamefont{A.}~\bibnamefont{Thess}}, \bibinfo{journal}{J.
  Fluid Mech.} \textbf{\bibinfo{volume}{350}}, \bibinfo{pages}{149}
  (\bibinfo{year}{1997}).

\bibitem[{\citenamefont{Boeck and Thess}(1998)}]{Boeck:1998}
\bibinfo{author}{\bibfnamefont{T.}~\bibnamefont{Boeck}} \bibnamefont{and}
  \bibinfo{author}{\bibfnamefont{A.}~\bibnamefont{Thess}},
  \bibinfo{journal}{Phys. Rev. Lett.} \textbf{\bibinfo{volume}{80}},
  \bibinfo{pages}{1216} (\bibinfo{year}{1998}).

\bibitem[{\citenamefont{Boeck and Thess}(1999)}]{Boeck:1999}
\bibinfo{author}{\bibfnamefont{T.}~\bibnamefont{Boeck}} \bibnamefont{and}
  \bibinfo{author}{\bibfnamefont{A.}~\bibnamefont{Thess}}, \bibinfo{journal}{J.
  Fluid Mech.} \textbf{\bibinfo{volume}{399}}, \bibinfo{pages}{251}
  (\bibinfo{year}{1999}).

\bibitem[{\citenamefont{Proctor}(1977)}]{Proctor:1977}
\bibinfo{author}{\bibfnamefont{M.~R.~E.} \bibnamefont{Proctor}},
  \bibinfo{journal}{J. Fluid Mech.} \textbf{\bibinfo{volume}{82}},
  \bibinfo{pages}{97} (\bibinfo{year}{1977}).

\bibitem[{\citenamefont{Busse and Clever}(1981)}]{Busse:Clever:1981}
\bibinfo{author}{\bibfnamefont{F.~H.} \bibnamefont{Busse}} \bibnamefont{and}
  \bibinfo{author}{\bibfnamefont{R.~M.} \bibnamefont{Clever}},
  \bibinfo{journal}{J. Fluid Mech.} \textbf{\bibinfo{volume}{102}},
  \bibinfo{pages}{75} (\bibinfo{year}{1981}).

\bibitem[{\citenamefont{Thual}(1992)}]{Thual:1992}
\bibinfo{author}{\bibfnamefont{O.}~\bibnamefont{Thual}}, \bibinfo{journal}{J.
  Fluid Mech.} \textbf{\bibinfo{volume}{240}}, \bibinfo{pages}{229}
  (\bibinfo{year}{1992}).

\bibitem[{\citenamefont{Chiffaudel et~al.}(1987)\citenamefont{Chiffaudel,
  Fauve, and Perrin}}]{Chiffaudel:1987}
\bibinfo{author}{\bibfnamefont{A.}~\bibnamefont{Chiffaudel}},
  \bibinfo{author}{\bibfnamefont{S.}~\bibnamefont{Fauve}}, \bibnamefont{and}
  \bibinfo{author}{\bibfnamefont{B.}~\bibnamefont{Perrin}},
  \bibinfo{journal}{Europhys. Lett.} \textbf{\bibinfo{volume}{4}},
  \bibinfo{pages}{555} (\bibinfo{year}{1987}).

\bibitem[{\citenamefont{Kek and M{\"u}ller}(1993)}]{Kek:1993}
\bibinfo{author}{\bibfnamefont{V.}~\bibnamefont{Kek}} \bibnamefont{and}
  \bibinfo{author}{\bibfnamefont{U.}~\bibnamefont{M{\"u}ller}},
  \bibinfo{journal}{Int. J. Heat Mass Transfer} \textbf{\bibinfo{volume}{36}},
  \bibinfo{pages}{2795} (\bibinfo{year}{1993}).

\bibitem[{\citenamefont{Nield}(1966)}]{Nield:1966}
\bibinfo{author}{\bibfnamefont{D.~A.} \bibnamefont{Nield}},
  \bibinfo{journal}{Z. Angew. Math. Phys.} \textbf{\bibinfo{volume}{17}},
  \bibinfo{pages}{131} (\bibinfo{year}{1966}).

\bibitem[{\citenamefont{Wilson}(1993)}]{Wilson:1993}
\bibinfo{author}{\bibfnamefont{S.~K.} \bibnamefont{Wilson}},
  \bibinfo{journal}{Journal of Engineering Mathematics}
  \textbf{\bibinfo{volume}{27}}, \bibinfo{pages}{161} (\bibinfo{year}{1993}).

\bibitem[{\citenamefont{Wilson}(1994)}]{Wilson:1994}
\bibinfo{author}{\bibfnamefont{S.~K.} \bibnamefont{Wilson}},
  \bibinfo{journal}{Phys. Fluids} \textbf{\bibinfo{volume}{6}},
  \bibinfo{pages}{3591} (\bibinfo{year}{1994}).

\bibitem[{\citenamefont{Thess and Nitschke}(1995)}]{Thess:Nitschke:1995}
\bibinfo{author}{\bibfnamefont{A.}~\bibnamefont{Thess}} \bibnamefont{and}
  \bibinfo{author}{\bibfnamefont{K.}~\bibnamefont{Nitschke}},
  \bibinfo{journal}{Phys. Fluids} \textbf{\bibinfo{volume}{7}},
  \bibinfo{pages}{1176} (\bibinfo{year}{1995}).

\bibitem[{\citenamefont{Hashim and Wilson}(1999)}]{Hashim:Wilson:1999}
\bibinfo{author}{\bibfnamefont{I.}~\bibnamefont{Hashim}} \bibnamefont{and}
  \bibinfo{author}{\bibfnamefont{S.~K.} \bibnamefont{Wilson}},
  \bibinfo{journal}{Int. J. Heat Mass Transfer} \textbf{\bibinfo{volume}{42}},
  \bibinfo{pages}{525} (\bibinfo{year}{1999}).

\bibitem[{\citenamefont{Thess and Nitschke}(18-22 November
  1991)}]{Thess:Nitschke:1992}
\bibinfo{author}{\bibfnamefont{A.}~\bibnamefont{Thess}} \bibnamefont{and}
  \bibinfo{author}{\bibfnamefont{K.}~\bibnamefont{Nitschke}}, in
  \emph{\bibinfo{booktitle}{Proceedings of the First European Symposium Fluids
  in Space}} (\bibinfo{address}{Ajaccio, France}, \bibinfo{year}{18-22 November
  1991}).

\bibitem[{\citenamefont{Miladinova and Slavtchev}(2001)}]{Miladinova:2001}
\bibinfo{author}{\bibfnamefont{S.~P.} \bibnamefont{Miladinova}}
  \bibnamefont{and} \bibinfo{author}{\bibfnamefont{S.~G.}
  \bibnamefont{Slavtchev}}, \bibinfo{journal}{Fluid Dynamics Research}
  \textbf{\bibinfo{volume}{28}}, \bibinfo{pages}{111} (\bibinfo{year}{2001}).

\bibitem[{\citenamefont{Davidson}(2001)}]{Davidson:2001}
\bibinfo{author}{\bibfnamefont{P.~A.} \bibnamefont{Davidson}},
  \emph{\bibinfo{title}{An Introduction to Magnetohydrodynamics}}
  (\bibinfo{publisher}{Cambridge University Press}, \bibinfo{year}{2001}).

\bibitem[{\citenamefont{Krasnov et~al.}(2004)\citenamefont{Krasnov, Zienicke,
  Zikanov, Boeck, and Thess}}]{Krasnov:2004}
\bibinfo{author}{\bibfnamefont{D.}~\bibnamefont{Krasnov}},
  \bibinfo{author}{\bibfnamefont{E.}~\bibnamefont{Zienicke}},
  \bibinfo{author}{\bibfnamefont{O.}~\bibnamefont{Zikanov}},
  \bibinfo{author}{\bibfnamefont{T.}~\bibnamefont{Boeck}}, \bibnamefont{and}
  \bibinfo{author}{\bibfnamefont{A.}~\bibnamefont{Thess}}, \bibinfo{journal}{J.
  Fluid Mech.} \textbf{\bibinfo{volume}{504}}, \bibinfo{pages}{183}
  (\bibinfo{year}{2004}).

\bibitem[{\citenamefont{Batchelor}(1956)}]{Batchelor:1956}
\bibinfo{author}{\bibfnamefont{G.~K.} \bibnamefont{Batchelor}},
  \bibinfo{journal}{J. Fluid Mech.} \textbf{\bibinfo{volume}{1}},
  \bibinfo{pages}{177} (\bibinfo{year}{1956}).

\bibitem[{\citenamefont{Boeck and Vitanov}(2002)}]{Boeck:2002}
\bibinfo{author}{\bibfnamefont{T.}~\bibnamefont{Boeck}} \bibnamefont{and}
  \bibinfo{author}{\bibfnamefont{N.}~\bibnamefont{Vitanov}},
  \bibinfo{journal}{Phys. Rev. E} \textbf{\bibinfo{volume}{65}},
  \bibinfo{pages}{037203} (\bibinfo{year}{2002}).

\end{thebibliography}

\end{document}